\documentclass[11pt]{iopart}

\usepackage{graphicx}
\expandafter\let\csname equation*\endcsname\relax

\expandafter\let\csname endequation*\endcsname\relax

\usepackage{mathtools}
\usepackage{microtype}
\usepackage[usenames,dvipsnames]{color}

\begin{document}



\title[MAST-U spectroscopy detachment]{Spectroscopic investigations of detachment on the MAST Upgrade Super-X divertor}



\author{K. Verhaegh$^{1}$, B. Lipschultz$^2$, J.R. Harrison$^1$, N. Osborne$^3$, A.C. Williams$^{1,2}$, P. Ryan$^1$, J. Allcock$^1$, J.G. Clark$^{1,3}$, F. Federici$^2$, B. Kool$^{4,5}$, T. Wijkamp$^{5,4}$, A. Fil$^1$, D. Moulton$^1$, O. Myatra$^{1,2}$, A. Thornton$^1$, T.O.S.J. Bosman$^{4,5}$, C. Bowman$^{2,1}$, G. Cunningham$^1$, B. P. Duval$^6$, S. Henderson$^1$, R. Scannell$^1$ and the MAST Upgrade team$^*$}
\address{$^1$ Culham Centre for Fusion Energy, Culham, United Kingdom} 
\address{$^2$ York Plasma Institute, University of York, United Kingdom}
\address{$^3$ University of Liverpool, Liverpool, United Kingdom}
\address{$^4$ Dutch Institute for Fundamental Energy Research DIFFER, Eindhoven, The Netherlands}
\address{$^5$ Eindhoven University of Technology, Eindhoven, The Netherlands}
\address{$^6$ Swiss Plasma Centre, \'{E}cole Polytechnique F\'{e}\'{e}rale de Lausanne, Lausanne, Switzerland}
\address{$^*$ See author list of “J. Harrison et al 2019 Nucl. Fusion 59 112011 (https://doi.org/10.1088/1741-4326/ab121c)}

\ead{kevin.verhaegh@ukaea.uk}

\begin{abstract}

We present the first analysis of the atomic and molecular processes at play during detachment in the MAST-U Super-X divertor using divertor spectroscopy data. Our analysis indicates detachment in the MAST-U Super-X divertor can be separated into four sequential phases: First, the ionisation region detaches from the target at detachment onset leaving a region of increased molecular densities downstream. The plasma interacts with these molecules, resulting in molecular ions ($D_2^+$ and/or $D_2^- \rightarrow D + D^-$) that further react with the plasma leading to Molecular Activated Recombination and Dissociation (MAR and MAD), which results in excited atoms and significant Balmer line emission. Second, the MAR region detaches from the target leaving a sub-eV temperature region downstream. Third, an onset of strong emission from electron-ion recombination (EIR) ensues. Finally, the electron density decays near the target, resulting in the bulk of the electron density moving upstream.

The analysis in this paper indicates that plasma-molecule interactions have a larger impact than previously reported and play a critical role in the intensity and interpretation of hydrogen atomic line emission characteristics on MAST-U. Furthermore, we find that the Fulcher band emission profile in the divertor can be used as a proxy for the ionisation region and may also be employed as a plasma temperature diagnostic for improving the separation of hydrogenic emission arising from electron-impact excitation and that from plasma-molecular interactions.

We provide evidences for the presence of low electron temperatures ($\ll 0.5$ eV) during detachment phases III-IV based on quantitative spectroscopy analysis, a Boltzmann relation of the high-n Balmer line transitions together with an analysis of the brightness of high-n Balmer lines. 

\end{abstract}


\section{Introduction}
\label{ch:introduction}

Power exhaust is a major challenge for the realisation of fusion energy as the expected un-mitigated target heat fluxes in a reactor can be significantly higher than the tolerable engineering limits at the target \cite{Loarte2007, Wenninger2014}. The heat load reductions that can be obtained by increased radiation alone (e.g. without momentum/particle dissipation) are limited to a factor 4-5 (not considering magnetic shaping) as they result in an increased ion target flux \cite{Verhaegh2021b}. The heat flux, arising from surface recombination of those ions, can exceed target engineering limits in reactor conditions if the upstream parallel energy flux density is sufficiently high. The target ion flux can be reduced through divertor detachment, during which plasma-atom/molecule interactions result in simultaneous power, particle and momentum dissipation in the divertor. Detachment is obtained when the target electron temperature reaches relatively low values ($T_t< \sim 5$ eV), which can be achieved by neutral recycling (e.g. higher upstream densities / fuelling) and/or extrinsic impurity seeding (increased radiative dissipation). 

Alternative Divertor Configurations (ADCs) aim to utilise variations in the divertor magnetic topology to enhance the power exhaust, compared to 'conventional' divertors, in three different ways. First, ADCs facilitate larger reductions of the parallel heat flow into the divertor before impinging on the target.  Secondly, ADCs reduce the detachment onset threshold (e.g. at lower core density or with lower impurity fractions) \cite{Lipschultz2016,Moulton2017}. Thirdly, ADCs can reduce the sensitivity of the displacement of the front-edge of the cold, detached, plasma (e.g. detachment 'front') to a 'physical' control parameter (e.g. density, upstream heat flux, impurity fraction) in steady-state \cite{Lipschultz2016,Havlickova2015,Moulton2017,Theiler2017,Valanju2009}. An increase of the detachment window and/or larger power dissipation in the divertor would allow the operation of future tokamak reactors with lower core impurity fractions and core radiation, which would benefit core, and hence fusion, performance. Drawbacks of ADCs in a reactor are however an increase in cost \cite{Militello2021}, an increase in engineering difficulty and, for the same vacuum vessel size, a reduction in plasma volume. Depending on the precise reactor parameters required, using ADCs may become essential \cite{Militello2021}.

The Mega Amp Spherical Tokamak - Upgrade (MAST-U) is a new spherical tokamak at the Culham Centre for Fusion Energy (CCFE) in the United Kingdom \cite{Morris2018,Fishpool2013, Harrison2019}. MAST-U was built specifically for investigating ADCs to 1) determine their feasibility and divertor performance; and 2) use magnetic shaping to improve our understanding of plasma-edge physics. MAST-U includes a tightly baffled lower and upper divertor chamber \cite{Fishpool2013} to facilitate a baffled double-null Super-X Divertor (SXD) \cite{Havlickova2015}. 

The SXD features an increased target radius resulting in a significantly increased 'total flux expansion', compared to the Conventional Divertor (CD). 'Total flux expansion' is defined as the ratio between the total magnetic field upstream (which we define to be at the x-point) and the total magnetic field at the target ($f_R = B^x / B^t$), analogously to \cite{Moulton2017}. Total flux expansion is different from poloidal flux expansion $f_x$, which is defined as the "ratio of the perpendicular flux surface spacing at the target and upstream" \cite{Theiler2017} $f_x = B_\theta^x B_\phi^t / B_\theta^t B_\phi^x$ where $B_\theta$ and $B_\phi$ denote the poloidal and toroidal components of the magnetic field and $x$, $t$ denotes upstream (which we define to be the x-point) and target.That is predicted by analytic and SOLPS-ITER models to reduce the detachment onset threshold \cite{Lipschultz2016,Moulton2017,Havlickova2015, Loarte2001,Petrie2013} in terms of upstream density and impurity fraction, increase the detachment window and reduce the detachment front sensitivity \cite{Lipschultz2016}. In contradiction to the predicted benefits of the Super-X divertor, TCV's (conventional aspect ratio tokamak) experimental results did, initially, not support such predictions \cite{Theiler2017} due to the openness of the, then non-baffled \cite{Theiler2017}, TCV divertor \cite{Fil2019submitted}. 

Together with the increased target radius and resultant total flux expansion, it is also important to trap the neutrals within the divertor region \cite{Fil2019submitted}. On MAST-U, this is achieved using tight baffling at the entrance to the divertor that increases neutral trapping \cite{Havlickova2015}. Keeping neutral trapping constant while varying the divertor configuration will test the benefit of increased total flux expansion \cite{Fil2019submitted}. The predictions of the beneficial effect of total flux expansion are in agreement with preliminary MAST-U results. During Ohmic density ramp discharges in the first MAST-U campaign ($P_{sep} \approx 550 kW$, $I_p = 650$ kA), the ion target flux roll-over point was observed to be at a $\sim 50\%$ lower Greenwald fraction for the Super-X than for the conventional divertor and strong heat flux reductions were observed in the SXD compared to the CD (by a factor $\sim 10$) \cite{Scannell2022, Thornton2022}. Due to this low detachment onset point and the lack of NBI injection, using the Super-X divertor resulted in detached divertor plasmas except for low core densities.

\subsection{This paper}

It is now critical to study the underlying physics processes in the Super-X divertor to obtain an understanding of 1) the physics mechanisms in the SXD; 2) how to diagnose the SXD; 3) whether existing models are sufficiently complete for simulating the SXD. Developing this understanding is a prerequisite for investigating alternative divertor configurations on MAST-U, as well as for ascertaining whether plasma-edge models include the processes necessary for simulating the MAST-U Super-X divertor correctly. Resolving gaps in modelling is critical for investigating ADCs in current day devices; which is needed to extrapolate to reactors. Investigating ADCs is an important risk mitigation strategy for DEMO \cite{Militello2021} and is of direct relevance of reactor-class devices with that employ an ADC, such as SPARC \cite{Kuang2020}, ARC \cite{Wigram2019} and STEP \cite{Wilson2020}.

This paper presents a first investigation of the plasma-atom and molecule interaction processes at play throughout the entire detached operational space of the MAST-U Super-X divertor (SXD). The level of detachment in the Super-X was separable into four phases. These start with a detachment of the ionisation region from the target and end with the electron density decaying near the target, resulting in the movement of the electron density bulk upstream. This is distinctively different from the conventional divertor operation in the first campaign where only the first phase of detachment progressive was observed as the density was scanned (up to $n_e/n_{GW} \sim 0.55$) and signs of MARFEs appeared in the core. The observed processes during detachment in the MAST-U Super-X divertor are not unique to MAST Upgrade. However, the precise sequence, location and evolution of the various processes are likely facilitated by the combination of 1) low divertor electron temperatures ($T_e \leq 0.2$ eV) with 2) low divertor electron densities ($n_e \approx 10^{19} m^{-3}$); which likely has been achieved due to the Super-X divertor although further research is required.

This phase progression is based upon line-of-sight spectroscopy through the Super-X divertor chamber. The 'intensity' (e.g. brightness in $ph/m^2/s$) of several hydrogen Balmer line intensities, together with the emission in the $D_2$ Fulcher band  between 595 to 615 nm, were monitored during detachment. Using the Balmer Spectroscopy Plasma-Molecular Interactions (BaSPMI) analysis developed previously \cite{Verhaegh2021a,Verhaegh2021,Verhaegh2021b,Verhaegh2019a}, the Balmer line brightnesses were decomposed into atomic and plasma-molecule interaction related emission component. In a relatively novel development, detailed analysis of the $n=9-20$ Balmer lines is used to provide evidence for low density ($n_e < 10^{19} m^{-3}$) and temperature conditions ($T_e \ll 0.5$ eV). 

Plasma-molecular interactions are shown to play a major role in the interpretation of hydrogen Balmer line measurements on MAST-U; consideration of such effects is critical in the interpretation of any MAST-U hydrogen spectroscopy and imaging data. This analysis required advancements in spectroscopic diagnostic analysis and its implications presented in this work are promising for aiding detachment control (section \ref{ch:relevanceFrontControl}), which will be important for fusion reactors. 

Quantitative evidence of the importance of plasma-molecular interactions in terms of spectroscopic interpretation as well as divertor physics is now available for TCV \cite{Verhaegh2021,Verhaegh2021a,Verhaegh2021b} as well as MAST-U (this work), with preliminary work also indicating the importance of these interactions in spectroscopic interpretations during deep detachment for JET \cite{Lomanowski2020,Karhunen2022,Karhunen2022a}. As such interactions are not fully included in plasma-edge modelling \cite{Verhaegh2021}, further investigating such reactions is important for reducing the uncertainties in the extrapolation of current day devices to future reactors - particularly for reactor concepts with ADCs. 

\section{MAST Upgrade overview}
\label{ch:mastu}  

\begin{figure}
    \centering
    \includegraphics[width=\linewidth]{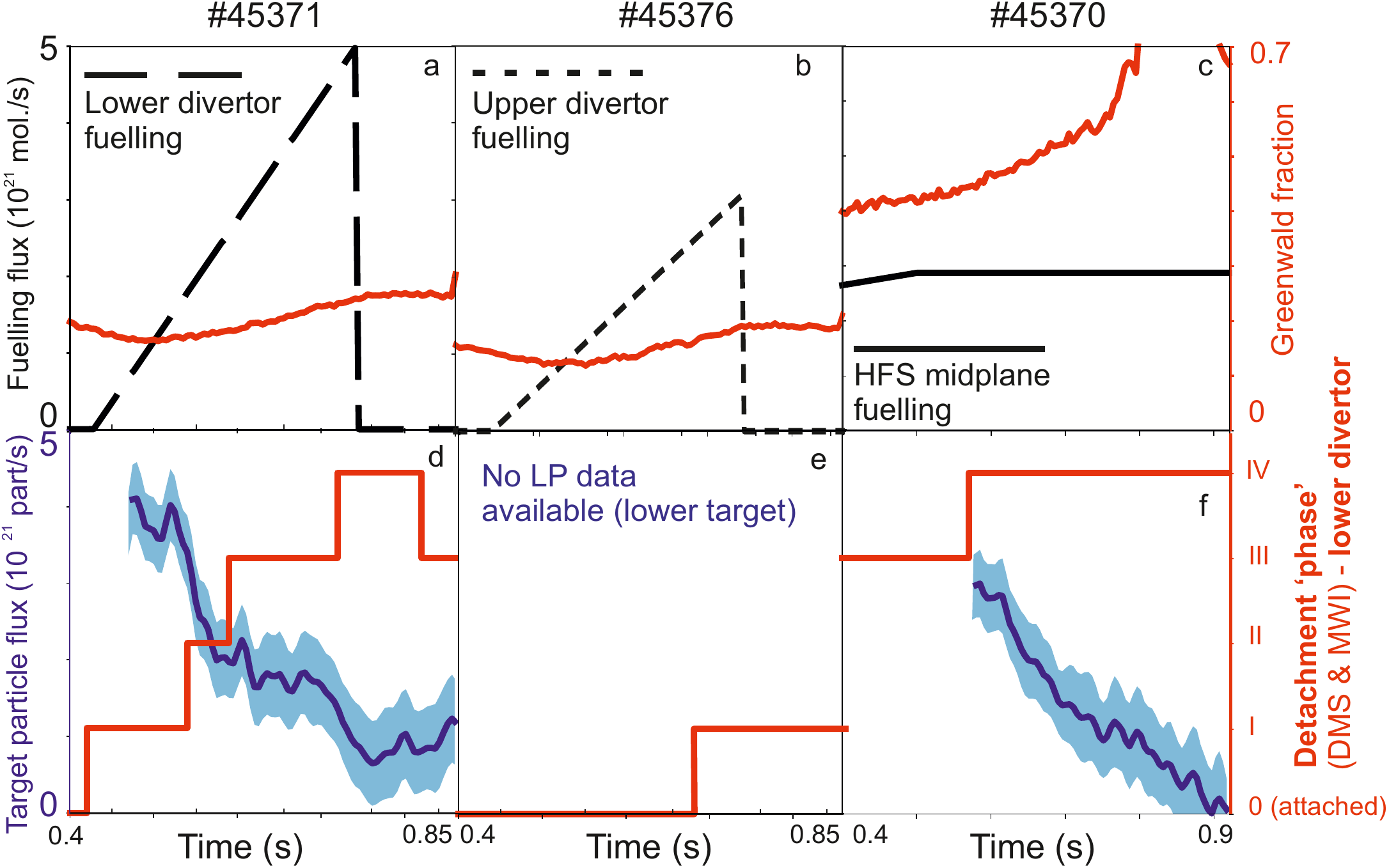}
    \caption{Overview of discharges \# 45371 (\emph{lower divertor} fuelled), \# 45376 (\emph{upper divertor} fuelled) and \# 45370 (high-field side (HFS) midplane fuelled). a,b) Core electron density and fuelling traces. c, d) \emph{Total lower outer target} integrated particle flux over 'tile 5' (figure \ref{fig:GeomLoS}), together with indicated divertor detachment phase based on \emph{lower divertor} spectroscopy (see section \ref{ch:results}). The ion target flux is only shown after the magnetic geometry of the Super-X divertor has been established (e.g. the strike point has been moved to 'tile 5' (see figure\ref{fig:GeomLoS}) and remains relatively steady) ($t=0.48$ s for \# 45371, 45376 and $t=0.58$ s for \# 45370).}
    \label{fig:DischargeOverview}
\end{figure}
\begin{figure}
    \centering
    \includegraphics[width=\linewidth]{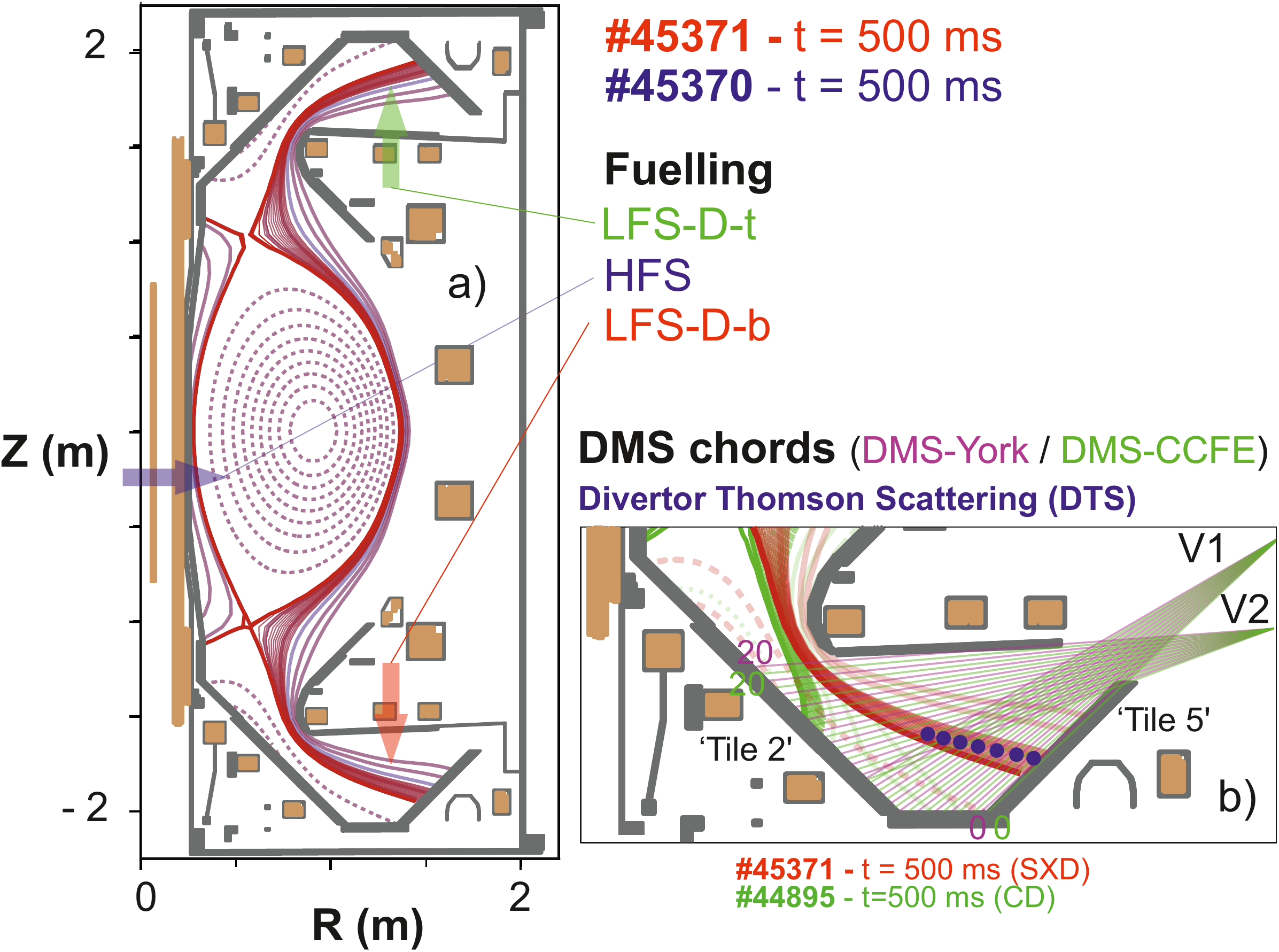}
    \caption{a) The magnetic geometries corresponding to the two discharges 
    \# 45371 (red) at 500 ms and \# 45370 (blue) at 600 ms, are shown together with the vessel geometry, poloidal field magnets and the fuelling valve locations utilised (`HFS', `LFS-D-b' and 'LFS-D-t'). b) Just the lower divertor region is shown along with the DMS spectroscopic chordal lines-of-sight originating from view points V1 and V2, which are both coupled to two spectrometers ('DMS-York' and 'DMS-CCFE') as well as the Divertor Thomson Scattering (DTS) measurement locations \cite{Clark2021}. The geometry of \# 45376 is not shown as it is a repeat of \# 45371 with different fuelling. In both a) and b), the Super-X separatrix strike point is incident on 'Tile 5'. A characteristic geometry for a conventional discharge has also been shown for comparison (\# 44895 at 500 ms), with the strike point incident on 'Tile 2', although this discharge is not used in the paper.}
    \label{fig:GeomLoS}
\end{figure}

Figure \ref{fig:DischargeOverview} shows fuelling and core density data together with the evolution of the \emph{total outer target} particle flux, as function of time for the main discharge used in this work (\# 45371, plasma current, $I_p = 650$ kA, $P_{sep} = 550 \pm 50$ kW, $P_{sep} B / R = 300$ kW T/m, $f_x$ = 7.2 (poloidal flux expansion), $f_R$ = 2.2 (total flux expansion) \footnote{\# 45376 is a repeat of \# 45371 and has the same parameters within uncertainties.}), and a discharge that is more deeply detached (\# 45370, $I_p = 450$ kA, $P_{sep} = 550 \pm 50$ kW, $P_{sep} B / R = 270 \pm 37$ kW/M, $f_x = 5.6$  (poloidal flux expansion), $f_R = 2.1$ (toroidal flux expansion), density (see figure \ref{fig:DischargeOverview}). \# 45370 and \# 45371 have slightly different magnetic equilibria as they are different discharge scenarios. Comparatively, a characteristic conventional divertor (strike point on 'tile 2' - see figure \ref{fig:GeomLoS}), such as \# 45474 has parameters: $I_p = 650$ kA, $P_{sep} = 550 \pm 100$ kW, $P_{sep} B / R = 300$ kW T / m, $f_R = 1.25$, $f_x = 3.9$. The level of detachment is delineated by the detachment phase given, based on spectroscopic analysis (see section \ref{ch:results}). After the divertor is formed, \# 45371 adds fuelling from the lower divertor valve (`LFS-D-b', shown in figure \ref{fig:GeomLoS} a), \# 45376 adds fuelling from the upper divertor valve (`LFS-D-t', shown in figure \ref{fig:GeomLoS} a) whereas \# 45370 employs the default midplane fuelling on the MAST-U inner wall (`HFS' - see figure \ref{fig:GeomLoS} a)). \footnote{Throughout the fuelling ramps, the divertor neutral pressure is expected to increase, although this measurement is unavailable during the first campaign. Therefore, we describe the observations during detachment in terms of the actuator: the fuelling strength; as we cannot delineate the different impacts of the core electron density and the divertor neutral pressure increase.} The values obtained for the \emph{total lower outer target} integrated particle flux are to be interpreted qualitatively at present, as some gaps remain in the Langmuir probe coverage at the outer target. Discharge \# 45376 is used in this work to provide a reference for an \emph{attached} discharge - injecting \emph{upper divertor} fuelling results in \emph{lower divertor} detachment at higher fuelling levels than \emph{lower divertor} fuelling (\# 45371). \footnote{These results indeed show that fuelling location influences the detachment evolution and determines which divertor detaches first, although investigating this in detail is beyond the scope of this paper.} Langmuir probe analysis for the lower divertor was not available for \# 45376. 

 Figure \ref{fig:GeomLoS} shows the magnetic geometry of both Super-X configurations with a strike point on 'Tile 5'. The poloidal distance between the target and the x-point is $\sim 110$ cm in both cases. As the X-point is significantly upstream of the divertor entrance, only the first 72 cm (poloidally) of the divertor leg can be monitored by diagnostics operating in the lower divertor chamber. For comparison, a conventional divertor geometry with a strike point on 'tile 2' is also shown, although this discharge is not used in this paper.

A new line-of-sight Divertor Monitoring Spectroscopy (DMS) system was developed and commissioned with lines of sight originating from two different lens-fibre arrangements - V1 and V2 (figure \ref{fig:GeomLoS} b). DMS uses two spectrometers in the lower divertor (`DMS-York' in purple; `DMS-CCFE' in green) with interleaved lines of sight to monitor two spectral regions simultaneously. For discharge \# 45371 \& \# 45376 we monitor: 1) the $n=5,6$ Balmer lines at medium spectral resolution (0.1 nm) and 2) part of the $D_2$ Fulcher band (595 - 615 nm) and $D\alpha$ (656 nm) simultaneously at low spectral resolution (0.4 nm). The second system includes a short-pass filter to attenuate the $D\alpha$ signal to better resemble that of the $D_2$ Fulcher band. For discharge \# 45370, DMS-York monitored the high-n ($n\geq 9$) Balmer lines simultaneously (365 - 385 nm) using a high spectral resolution (0.06 nm), whereas DMS-CCFE monitored the CIII (465 nm) and $D\beta$ Balmer line (486 nm) (not used in this work).  



\section{Spectroscopic MAST-U Super-X detachment observations and their interpretation}
\label{ch:results}

\begin{figure}
    \centering
    \includegraphics[width=0.9\linewidth]{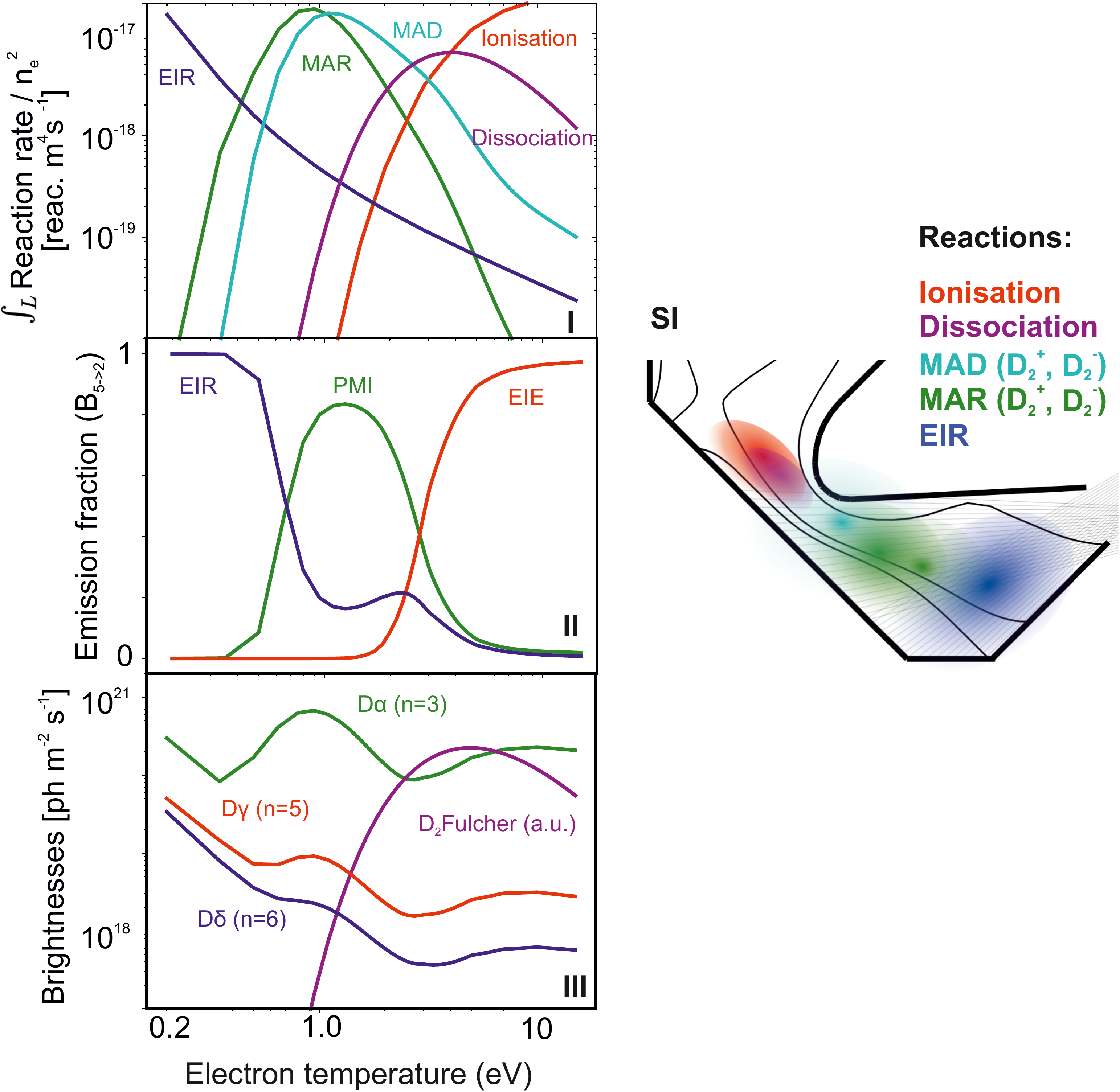}
    \caption{Illustration of difference and correspondence between reactions, emission processes and emission measurements. I, II, III) Simplified model results (section \ref{ch:emissmodel}, with MAST-U scaling laws assuming $n_e = 10^{19} m^{-3}$) as function of $T_e$ of the various reaction rates (I), emission fraction contributions to $D\alpha$ (II), total $D\alpha, D\gamma, D\delta$ Balmer line brightnesses and $D_2$ Fulcher band brightness (a.u.) (III). SI) Schematic illustration of the various reactions in the MAST-U divertor in 2D during deep detachment. The magnetic geometry in this illustration has been obtained from a SOLPS-ITER simulation (from \cite{Myatra}).}
    \label{fig:proc_schem}
\end{figure}

It is important to distinguish three concepts in this paper, illustrated in figure \ref{fig:proc_schem}. \begin{enumerate}
    \item \emph{Reactions:} Plasma-atom and molecular interactions lead to reactions that influence the ion and neutral atom particle balance, such as ionisation, Electron-Ion Recombination (EIR), electron-impact dissociation, Molecular Activated Ionisation (MAI), Molecular Activated Recombination (MAR) and Molecular Activated Dissociation (MAD). MAST-U characteristic line-integrated reaction rates are plotted as function of $T_e$ in figure \ref{fig:proc_schem} I, using the model from section \ref{ch:importance_plasmamol}. Figure \ref{fig:proc_schem} SI shows a schematic cartoon indicating the various processes in the MAST-U divertor during deep detachment.
    \item \emph{Emission processes:} Electron-impact excitation (EIE), EIR and Plasma-Molecule Interactions (PMI) generate excited atoms that emit hydrogenic atomic line emission.  Figure 
    \ref{fig:proc_schem} II shows MAST-U characteristic fractional emission contributions to the $D\alpha$ brightness as function of $T_e$ using the model from section \ref{ch:importance_plasmamol}.
    \item \emph{Emission measurements:} Any hydrogen atomic emission (e.g. Balmer line) is, in general, due to a combination of emission processes, which can vary as function of divertor conditions. Detailed analysis is often required to infer information on the dominant emission processes from hydrogenic atomic line measurements. Figure 
    \ref{fig:proc_schem} III shows MAST-U characteristic total brightnesses of $D\alpha$, $D\gamma$ and $D\delta$ as function of $T_e$ together with the $D_2$ Fulcher band emission brightness, using the model from section \ref{ch:importance_plasmamol}.
\end{enumerate} 

In this work, we show hydrogen atomic emission (e.g. Balmer line) measurements analysed using BaSPMI, to infer the quantitative emission processes contributing to the $D\alpha$ emission. Figure \ref{fig:proc_schem} shows this also provides qualitative information on 1) appearance; 2) location and 3) relative magnitude of the reactions active in the Super-X divertor and their evolution during detachment. Balmer emission from EIE provides information on the ionisation source; from EIR on the EIR ion sink and from plasma-molecular interactions (mostly $D_2^+, D^-$) on MAR and MAD. The $D_2$ Fulcher band emission, arising from electronically excited molecules after electron-impact collisions, is correlated with $D_2$ electron-impact dissociation. A quantitative analysis of the ion sources and sinks in the divertor remains outside of the scope of this paper.

In section \ref{ch:importance_plasmamol}, the simplified approach from figure \ref{fig:proc_schem} is explained in more detail and used for both TCV and MAST-U characteristic conditions, as well as extrapolated to higher electron densities ($10^{21} m^{-3}$) that are more reactor relevant. Increases to the electron density will increase the EIR emission contribution and its reaction rate greatly (with $n_e^{2-3}$), which are discussed in section \ref{ch:importance_plasmamol}. However, even in such conditions, the discussed relation between reaction rates, hydrogen emission contributions and the total hydrogen emission remains intact.

\begin{figure}
    \centering
    \includegraphics[width=\linewidth]{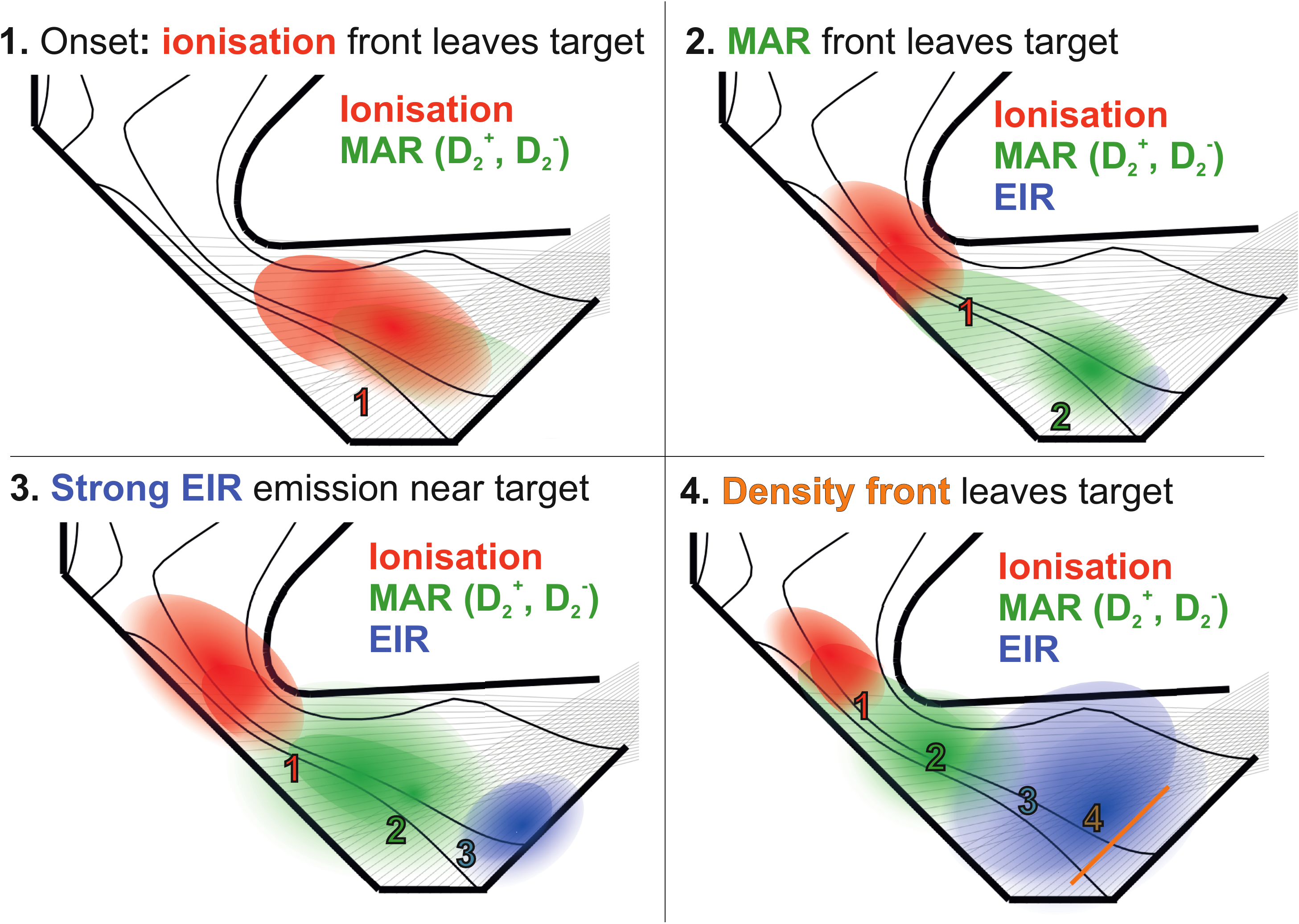}
    \caption{Schematic overview of the four inferred MAST-U Super-X detachment phases in terms of the reactions occurring in the divertor. Also shown is the Super-X plasma geometry and the DMS spectroscopic viewing chords. The numbers shown indicate: (1) the back-end of the ionisation region; (2) the back-end of the Molecular Activated Recombination (MAR) region; (3) the front-end of the Electron Ion Recombination (EIR) region; and (4) the back-end of the electron ion recombination / density region.  The magnetic geometry in this illustration has been obtained from a SOLPS-ITER simulation (from \cite{Myatra}).}
    \label{fig:detach_schem}
\end{figure}

The DMS data analysis indicates that detachment in the MAST-U Super-X divertor can be separated in four different phases (figure \ref{fig:detach_schem}), which has been generally observed during Super-X divertor operation in the first MAST-U campaign. 

\begin{enumerate}
    \item Detachment onset: the ionisation front detaches from the target and moves towards the divertor entrance, leaving behind a region where plasma-molecular interactions involving molecular ions generate excited atoms composing a majority of the Balmer line emission (section \ref{ch:ionisation_movement}). These molecular ions include $D_2^+$ (likely the dominant contributor \cite{Verhaegh2021b}) created from molecular charge exchange ($D^+ + D_2 \rightarrow D_2^+ + D$)
    \footnote{Non-dissociative ionisation $e^- + D_2 \rightarrow D_2^+ + 2 e^-$ occurs at higher temperatures in the ionisation region.} and possibly $D^-$ ($D_2^- \rightarrow D^- + D$).
    \item The peak in Balmer line emission associated with molecular ions detaches from the target; indicative of an upstream movement of the MAR and MAD regions \cite{Verhaegh2021b}. The MAR \& MAD rates near the target are reduced as the plasma temperature drops below 1 eV, inhibiting the capability of creating molecular ions (section \ref{ch:MAR_movement}, figure \ref{fig:proc_schem} I, II).
    \item Appearance of strong electron-ion recombination (EIR) emission near the target (section \ref{ch:strong_EIR}) and  evidence for $T_e \ll 0.5$ eV (section \ref{ch:lowTe}). 
    \item The EIR related emission region detaches from the target and moves towards the divertor entrance; this suggests that the electron density bulk moves completely away from the target.
\end{enumerate}

\begin{figure}
    \centering
    \includegraphics[width=0.85\linewidth]{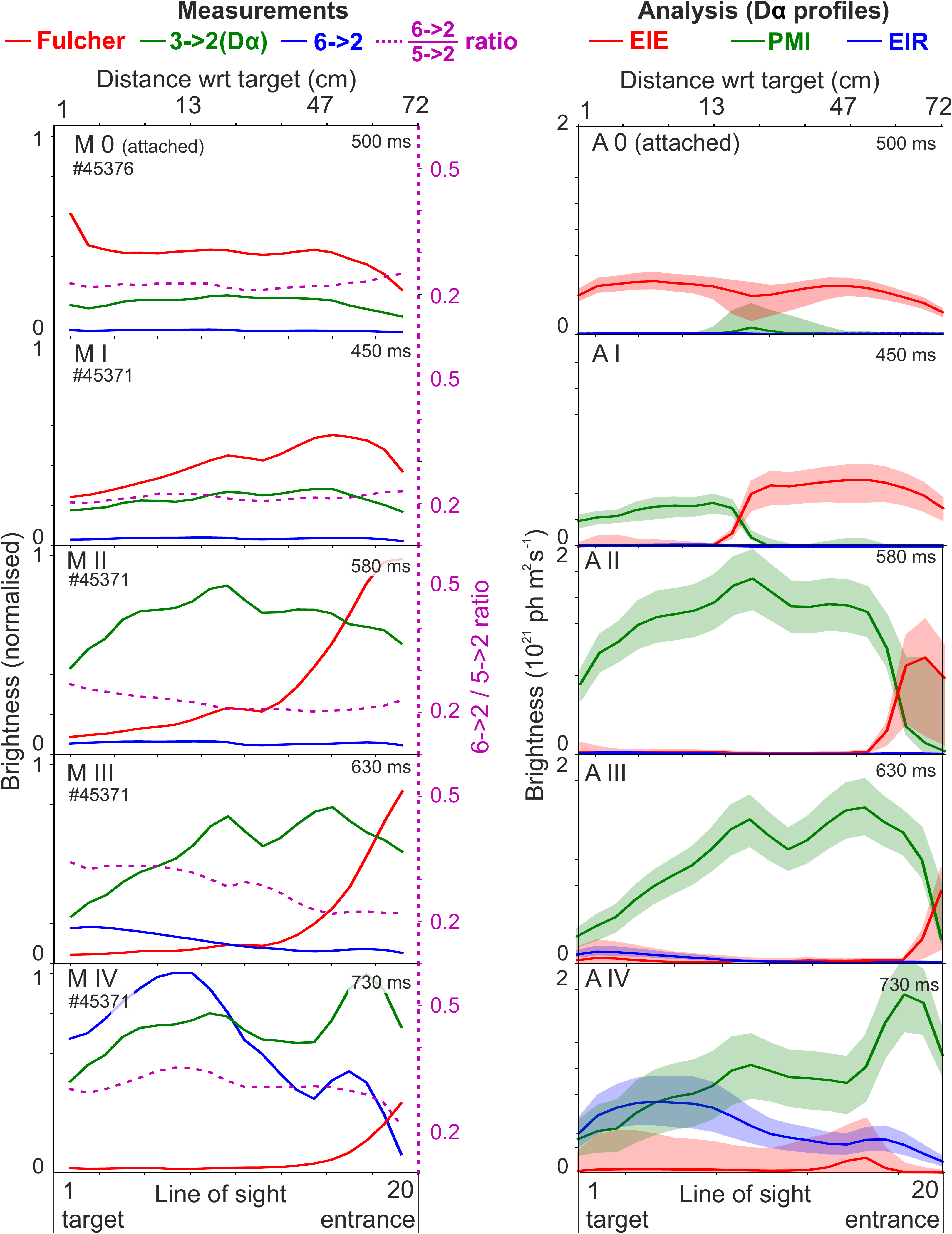}
    \caption{(M 0) through (M IV): Measured and normalised brightness profiles in the lower divertor during MAST-U pulse \# 45376 (M 0 - attached) and \# 45371 as function of time (M I - M IV) for the Fulcher band (595-615 nm), $n=3,6$ Balmer lines as well as the $6\rightarrow2 / 5\rightarrow2$ Balmer line ratio along the various spectroscopic chords at four different times, spanning the attached regime (0) and the four phases of detachment (I-IV). A 0) through A IV): Analysed BaSPMI  \cite{Verhaegh2019a,Verhaegh2021} from measurements M 0 (\# 45376 - attached reference) and \#45371 as function of time (M I - M IV) showing the brightness profiles of the $D\alpha$ associated with electron-impact excitation ("EIE" of $D$), Plasma-Molecule Interactions ("PMI" of $D_2, D_2^+, D^-$) and electron-ion recombination ("EIR" of $D^+$). The indicative poloidal distance, along the divertor leg, between the target and intersection of the line-of-sight with the separatrix are shown on top. The 'EIE', 'PMI' and 'EIR' emission regions are indicative of the 'ionisation', 'MAR \& MAD' and 'EIR' regions, respectively \cite{Verhaegh2021b}.}
    \label{fig:EmissionTrends}
\end{figure}

Figure \ref{fig:EmissionTrends} M0-IV shows Balmer line emission intensities, ratios and the $D_2$ Fulcher band emission intensity as a profile along the DMS lines of sight during the attached (0) phase for discharge \# 45376 and four identified different detachment phases (I-IV) for discharge \# 45371. Separations of the different $D\alpha$ emission processes, using BaSPMI analysis, accompany these measurements in figure \ref{fig:EmissionTrends} AI -AIV. BaSPMI application is further discussed in \ref{ch:baspmi} and is independently verified using a new Bayesian approach (\ref{ch:bayspmi}). The observations and sequence of the four different detachment phases have been observed over a range of plasma currents, variations in super-X geometry (e.g.different poloidal flux expansion and strike point locations) and fuelling locations (upper divertor, lower divertor, high field side, low field side). Estimates of electron density and temperature from both spectroscopy and divertor Thomson scattering are compared for \# 45371 in section \ref{ch:DMS_DTS_Compa}.

\subsection{Detachment phase 1: ionisation movement and MAR appearance}
\label{ch:ionisation_movement}

Before the detachment onset \footnote{Since \# 45371 is never fully detached in the lower divertor, the data shown has been obtained from \# 45376 at 500 ms, which is the same scenario as \# 45371 but employs upper divertor fuelling.}, we generally observe: 1) line ratios indicative of EIE, 2) a roughly flat profile of the $n=3,5,6$ Balmer line intensities as well as the Fulcher band emission (with slight peaking near the target) along the divertor leg (figure \ref{fig:EmissionTrends} M0). BaSPMI analysis indicates a flat profile of EIE along the divertor leg, suggestive of an attached ionisation front. This is similar to TCV observations \cite{Verhaegh2017,Verhaegh2019,Verhaegh2021b}, where a strong correlation between the Fulcher band emission and the ionisation region were found \cite{Verhaegh2021b,Verhaegh2021}. 

At the start of \# 45371, near the detachment onset, the Fulcher emission region detaches from the target (figure \ref{fig:EmissionTrends} MI). Simultaneously, both the Balmer line emission and their ratios remain constant over the entire divertor leg (Figure \ref{fig:EmissionTrends} MI), \emph{similar to the attached phase (figure \ref{fig:EmissionTrends} M0)}.

As Balmer line emission ratios associated with PMI and EIE are similar \cite{Verhaegh2021} (section \ref{ch:importance_plasmamol}), \emph{the combined observations of the Fulcher emission region, Balmer line ratio and Balmer line intensity are key to show how neglecting excited atoms from PMI can lead to incorrect interpretations} (figure \ref{fig:EmissionTrends} MI): the ionisation region spans the entire divertor leg and remains attached at the target (similar to figure \ref{fig:EmissionTrends} MI). However, this would be inconsistent with the Fulcher emission that suggests that the region near the target is too cold for electron-impact molecular dissociation (and thus too cold to ionise atoms).

\emph{Therefore, we hypothesise that the Fulcher band emission brightness profile may be employed to separate hydrogenic emission from PMI and EIE;} which is an important tool for hydrogen Balmer line analysis. Although the Balmer line trends in figure \ref{fig:EmissionTrends} M0 and MI are almost identical, \emph{the $D_2$ Fulcher band brightness profile along the divertor leg is different}. Employing the Fulcher band as a filter implies that a movement of the \emph{cold-end of the Fulcher emission}, suggests movement of the region where the plasma is too cold to ionise and thus \emph{the ionisation front}. Likewise, an \emph{absence of Fulcher emission} near the target, suggests - through exclusion - that EIR or PMI are the dominant hydrogenic emission processes near the target. Balmer line ratios can then be used to distinguish between EIR and PMI.   

This hypothesis is analysed more quantitatively in section \ref{ch:fulcher_ionisation}, where it is shown that the Fulcher emission profile may be employed as a quantitative temperature constraint. That constraint is employed in our BaSPMI analysis (described in \ref{ch:baspmi}), which facilitates separating EIE and PMI related emission where the Fulcher emission is low and thus the Fulcher information constraints the temperature to be relatively low \footnote{Fulcher $T_e$ estimates are shown in figure \ref{fig:neTeDMSDTS}}. As the fuelling is increased, the EIE region moves quickly from near the target to near the divertor entrance (indicative of an ionisation front movement), leaving a region where Balmer line emission is dominantly from PMI (indicative of a growing region with significant MAR \& MAD). The magnitude of MAR \& MAD related emission increases over time in detachment phase I and its magnitude reaches sufficiently large values that it is expected that the MAR ion sinks are significant.  



\subsection{Detachment phase 2: movement of MAR region and obtaining sub-eV temperatures}
\label{ch:MAR_movement}

As the fuelling is further increased, the Fulcher emission moves further towards the divertor entrance. Below the Fulcher emission region, the $D\alpha$ emission increases whilst the $6\rightarrow2/5\rightarrow2$ Balmer line ratio remains at levels expected of a EIE or PMI dominant plasma (figure \ref{fig:EmissionTrends} MII). This indicates the $D\alpha$ emission is predominantly from PMI, which is consistent with BaSPMI analysis (figure \ref{fig:EmissionTrends} AII). Ultimately, the $D\alpha$ (PMI) emission region starts to move from the target towards the divertor entrance, figure \ref{fig:EmissionTrends} MII \& AII; which is suggestive of a movement of the MAR \& MAD regions.

Plasma-molecular interactions leading to hydrogen emission and MAR/MAD involve $D_2^+$ and/or $D^-$. Creating those species below 5 eV requires the presence of $D_2 (\nu)$ molecules that are excited to relatively high vibrational levels \cite{Reiter2018,Verhaegh2021b,Holm2022}. Below a certain temperature,  both the probability of exciting molecules vibrationally drops \cite{Holm2022} and the ions have insufficient energy to promote molecular charge exchange \cite{AMJUEL,Reiter2008,Ichihara2000}, reducing the $D_2^+/D_2$ ratio \cite{Verhaegh2021b,AMJUEL,Sawada1995,Ichihara2000} for decreasing temperature. The intensity of MAR/MAD with decreasing temperature (and its associated hydrogen emission) is a competition between two processes: 1) an increase of the $D_2$ density; 2)\ a decrease in the $D_2^+/D_2$ ratio. As a result, the hydrogen emission associated with MAR and MAD is expected to occur between 0.5 ('cold-end' of the emission region) and 3.5 eV ('hot-end' of the emission region) and to peak $\sim 0.8$ eV (section \ref{ch:importance_plasmamol}). 

As the fuelling increases further, the cold-end of the MAR region moves increasingly further from the target until it reaches the divertor entrance. This would suggest that the majority of the divertor chamber has achieved sub-eV electron temperatures.


\subsection{Detachment phase 3: appearance of strong signs of electron-recombination}
\label{ch:strong_EIR}

As fuelling is further increased, the Fulcher emission region moves into the divertor throat, out of spectroscopy view. The MAR emission region moves towards the divertor throat and, ultimately, a strong increase in both $n=6$ Balmer line intensity and the $n=6/5$ Balmer line intensity ratios are observed near the target, figure \ref{fig:EmissionTrends} MIII. This implies a transition from PMI dominated emission to EIR dominated emission near the target, which is consistent with our BaSPMI analysis (figure \ref{fig:EmissionTrends} A III). EIR preferentially contributes to high- and medium-n Balmer line emission: BaSPMI analysis at 630 ms indicates the EIR contribution near the target to $n=5 (6)$ is 80-90\% (90-95 \%), although this is smaller for $D\alpha$ . 

\begin{figure}
    \centering
    \includegraphics[width=0.9\linewidth]{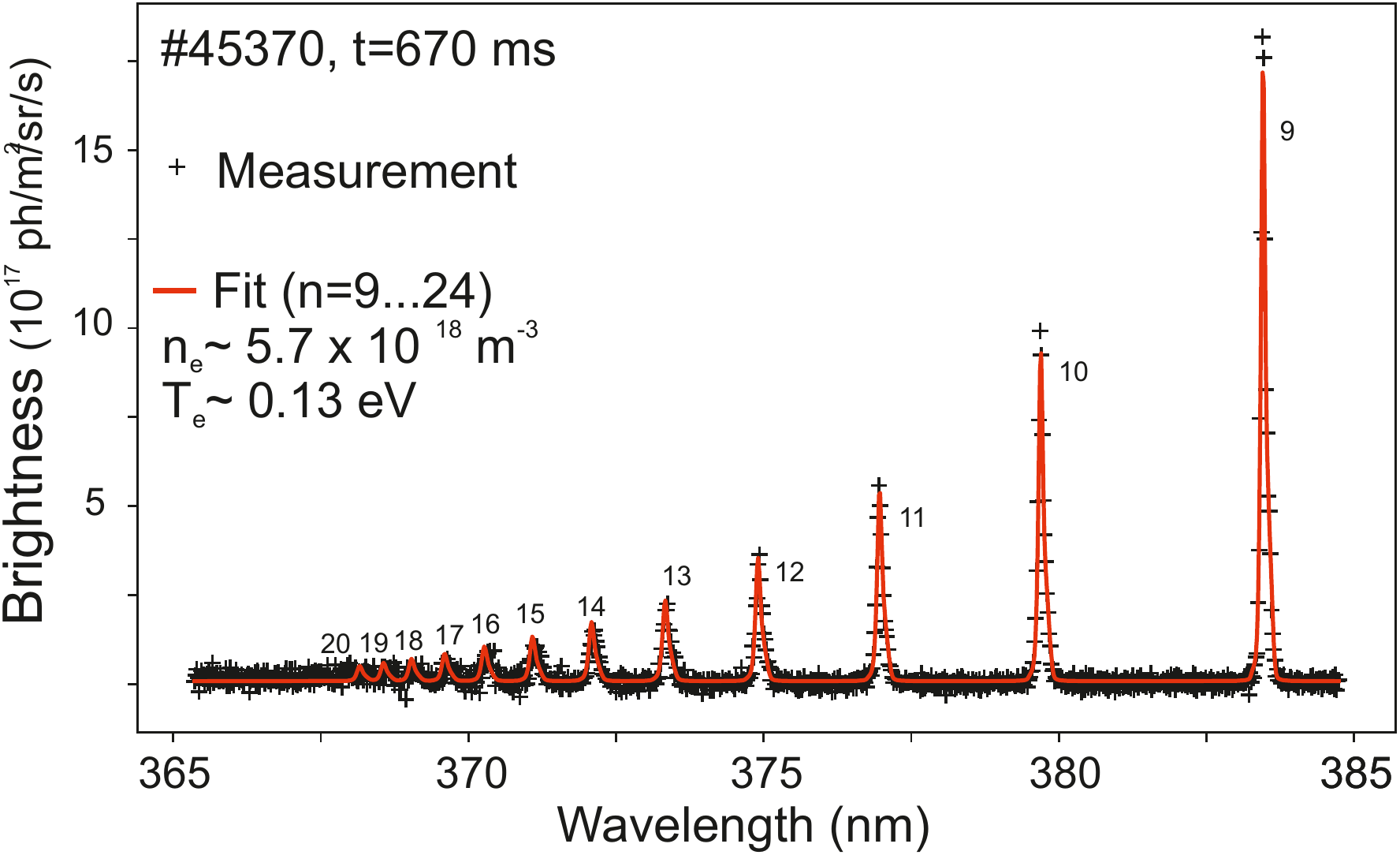}
    \caption{Measured high-n Balmer line spectra for line of sight \# 10 (midway divertor) at 670 ms for \# 45370, together with a high-n Balmer line fit (red). $n_e$ and $T_e$ (using a Boltzmann distribution) are inferred from the high-n Balmer line fit, indicating a low electron density/temperature regime. The $n=9$ brightness in this case is $4 \times 10^{18} ph m^{-2} s^{-1}$.}
    \label{fig:HighNFit}
\end{figure}

The entire high-n Balmer line spectra (discharge \# 45370) has been fitted using a Stark broadening model \cite{Verhaegh2019a,Lomanowski2015} together with a Boltzmann distribution to model the ratio between the intensity of the various high-n Balmer lines \cite{Lipschultz1998,Terry1998} as function of $T_e$. Figure \ref{fig:HighNFit} shows a representative example of the fit obtained. Similar analysis results have been obtained over many discharges, including a range of plasma currents.

Our fit results (figure \ref{fig:HighNFit}) indicate that the EIR emission along the DMS viewing chords originates from a region with low electron density ($n_e < 1 \times 10^{19} m^{-3}$) and temperature ($\ll 0.5$ eV - ($0.1-0.2$ eV)). These temperatures and densities can be considered as EIR emission-weighted quantities along the line of sight \cite{Verhaegh2019,Verhaegh2021} that, as such, may not correspond to the separatrix density/temperature. These low electron density estimates are consistent with the Inglis-Teller limit \cite{Kunze2009}: by averaging over 15 acquisition frames ($\sim 200 ms$) up to $n=21$ Balmer line is resolved, for which the Inglis-Teller limit suggests that $n_e < 1.6 \times 10^{19} m^{-3}$ \footnote{It is unclear whether the $n=21$ observation limit is due to a limited signal to noise ratio or due to a merging of the continuum. As such, the Inglis-Teller limit can only be used to get an upper limit electron density estimate in this case.}. The relatively low electron densities in the MAST-U divertor are likely the reason why signs of EIR are only present when the electron divertor temperatures are very low ($\ll 0.5$ eV) and is one of the reasons why the detachment of the PMI emission region occurs before signs of EIR are present. The reason of such low electron densities is currently unknown and requires further investigation.

\subsubsection{Additional proof for $T_e < 0.5$ eV}
\label{ch:lowTe}

The obtained electron temperatures from the Boltzmann fit (detachment phase III-IV) are between 0.1 - 0.2 eV. To support these electron temperature estimates ($T_e \ll 0.5$ eV), we provide four additional sources of evidence in this subsection: 1) an analysis of the required path-lengths needed to model the high-n Balmer line brightness; 2) BaSPMI analysis; 3) comparison against simplified emission modelling. 4) Those results are consistent with divertor Thomson scattering, which indicates $T_e < 0.5$ eV during deep detachment phases III and IV (see section \ref{ch:DMS_DTS_Compa}).

First, we perform a consistency check of $T_e$ using the monitored high-n Balmer line emission brightness. For a given electron density ($n_e$), the monitored brightness depends on the size of the emission region along the line of sight ($\Delta L$) and the electron temperature ($T_e$) - equation \ref{eq:recomb_B}, assuming that all high-n Balmer line emission ($n\geq9$) is due to EIR - which is verified in \ref{ch:highnEIR}. Here, $PEC_{n\geq9\rightarrow2}^{rec} (n_e, T_e)$ is the recombination emission coefficients for high-n Balmer lines that depends on $n_e$ and increases exponentially with linearly decreasing $T_e$. For a known $n_e$, $\Delta L$ decreases with decreasing $T_e$. If we assume $T_e = 0.2$ eV (ADAS default temperature limit \cite{OMullane}), one would require a path-length of 1.6 m near the target and 0.2 m near the divertor entrance at $t=600$ ms of \# 45370. MAST-U features a Multi-Wavelength Imaging (MWI) diagnostic \cite{Feng2021}, which can obtain spectrally filtered images of the divertor plasma, similar to MANTIS on TCV \cite{Perek2019submitted,Perek2021,Perek2022}. Inverting the MWI channel that hosts a filter for the $n=9$ Balmer line, $\Delta L$ would limited to 0.20-0.40 m physically, this provides further evidence for a $T_e = 0.1 - 0.2$ eV regime obtained from the Boltzmann fit. 

\begin{equation}
\begin{aligned}
    B_{n\geq9\rightarrow 2} &\approx \Delta L n_e^2 PEC_{n\geq9\rightarrow2}^{rec} (n_e, T_e) \\
    \Delta L &\approx \frac{B_{n\geq9\rightarrow 2}}{ n_e^2 PEC_{n\geq9\rightarrow2}^{rec} (n_e, T_e)}
    \end{aligned}
    \label{eq:recomb_B}
\end{equation}

Secondly, BaSPMI analysis (\# 45371) provides additional evidence for $T_e \ll 0.2$ eV. BaSPMI estimates the recombination rate and recombinative temperature $T_e^R$ by analysing the intensity of the EIR component of the medium-n Balmer line brightness \cite{Verhaegh2019a}, $B_{n\rightarrow2}^{rec}$. Lower $T_e^R$ are obtained for higher $B_{n\rightarrow2}^{rec}$ until a maximum reference value for $B_{n\rightarrow2}^{rec}$ is reached at the $T_e^R = 0.2$ ADAS limit. BaSPMI analysis for \# 45371 indicates that during detachment phases III-IV (figure \ref{fig:EmissionTrends} A III and A IV) a large portion of the Monte Carlo inferred values for $B_{6\rightarrow2}^{rec}$ exceeds their maximum reference values for $B_{6\rightarrow2}^{rec}$; providing evidence that $T_e$ may be smaller than 0.2 eV. 

Thirdly, the simplified emission modelling as function of $T_e$ in figure \ref{fig:proc_schem} (explained in section \ref{ch:importance_plasmamol}), leads to a predicted maximum $n=6$ Balmer line emission brightness at $T_e=0.2$ eV (ADAS limit) of $\sim 3 \cdot 10^{19} ph/m^2/s$. The peak $n=6$ emission brightness measured in the EIR region of \# 45371 (detachment phase IV) is, however, larger $\sim 5 \cdot 10^{19} ph/m^2/s$ (figure \ref{fig:EmissionTrends} MIV). Although the difference between these two values is not that large, the electron density assumed in figure \ref{fig:proc_schem} is $n_e = 10^{19} m^{-3}$, which is higher than obtained in the experiment in this regime according to divertor Thomson scattering and high-n ($n \geq 9$) Balmer line analysis. Therefore, this comparison provides some additional evidence of $T_e \leq 0.2$ eV.


\subsection{Detachment phase 4: EIR emission movement and bulk density displacement}
\label{ch:density_front_movement}

\begin{figure}
    \centering
    \includegraphics[width=0.5\linewidth]{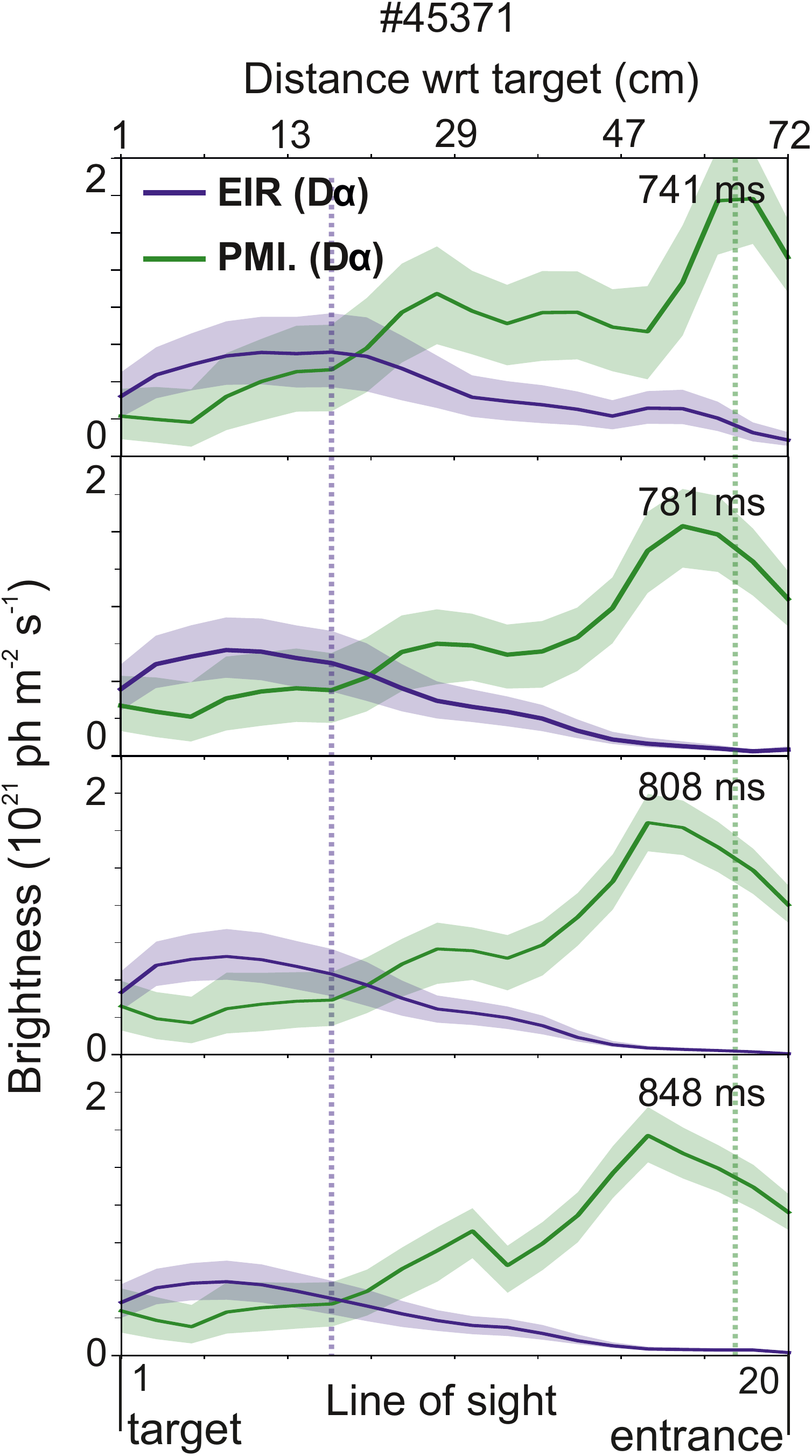}
    \caption{Measurements of the movement of the MAR and EIR emission regions based on BaSPMI analysis before the fuelling cut (750 ms) and three time frames after the fuelling cut. As a guide, dotted lines are shown corresponding to the peak in MAR (green) and EIR (blue) before the fuelling cut.}
    \label{fig:GasCut}
\end{figure}

We have shown that through phase III, increased fuelling changes the Fulcher emission intensity such that it becomes almost absent from the divertor chamber, the MAR peak reaches the divertor entrance with the EIR region further expanding towards the divertor entrance. Ultimately, in phase IV, the cold-end of the EIR emission region leaves the target (figure \ref{fig:EmissionTrends} IV). As mentioned in the previous section, here, the dominant Balmer line emission process is EIR (figure \ref{fig:EmissionTrends} AIV). Since the temperature in the divertor must decrease (or stay constant) from upstream to downstream, a decrease in the EIR emission must correspond to a $n_e$ decrease near the target. As fuelling is increased, the cold-end of the EIR emission region moves further towards the divertor entrance, indicating a significant reduction of the target electron densities and thus a bulk electron density displacement. 

100 ms before the final $I_p$ ramp-down in \# 45371 (figure \ref{fig:DischargeOverview} a, time of 750 ms), we stop all fuelling and observe how the divertor plasma responds. As a result, the discharge transitions back to detachment phase III after 50 ms, as shown from the EIR and PMI $D\alpha$ emission in figure \ref{fig:GasCut}: the peak in MAR emission moves back into the divertor chamber and the cold-end of the EIR emission region moves back towards the target.

\begin{figure}
    \centering
    \includegraphics[width=\linewidth]{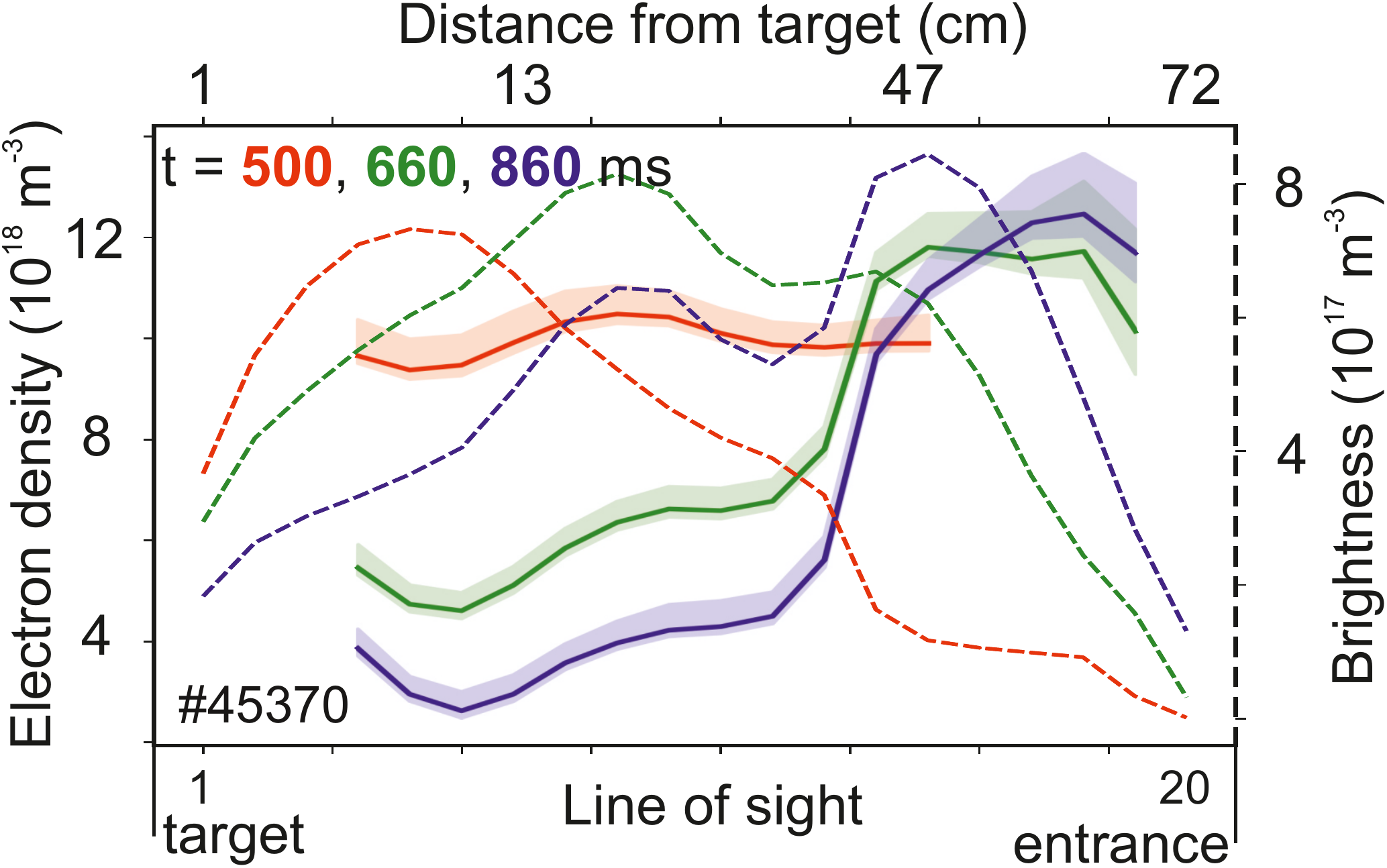}
    \caption{Spatial profiles, from near the target to the divertor baffle, of $n \geq 9$ (summed) Balmer line brightness (dotted lines) as well as inferred electron densities from Stark broadening (solid lines) at 500 (red), 670 (green) and 860 (blue) ms of \# 45370.}
    \label{fig:DenFrontMovement}
\end{figure}

Discharge \# 45370, which is more deeply detached and ultimately terminates in a disruption engendered by a MARFE, has high-n Balmer line coverage, permitting higher accuracy $n_e$ inferences (see figure \ref{fig:HighNFit}). Figure \ref{fig:DenFrontMovement} shows the movement of the $n=9$ Balmer line emission, towards the divertor entrance, as the fuelling is increased. This is correlated with with a sharp decay in the inferred electron density near the target, leading to the electron density bulk moving upstream as is illustrated by the measured electron density profiles at 660 and 860 ms in the lower divertor (figure \ref{fig:DenFrontMovement}). 

\subsection{Comparison between divertor Thomson scattering and spectroscopically inferred electron densities and temperatures}
\label{ch:DMS_DTS_Compa}

\begin{figure}
    \centering
    \includegraphics[width=\linewidth]{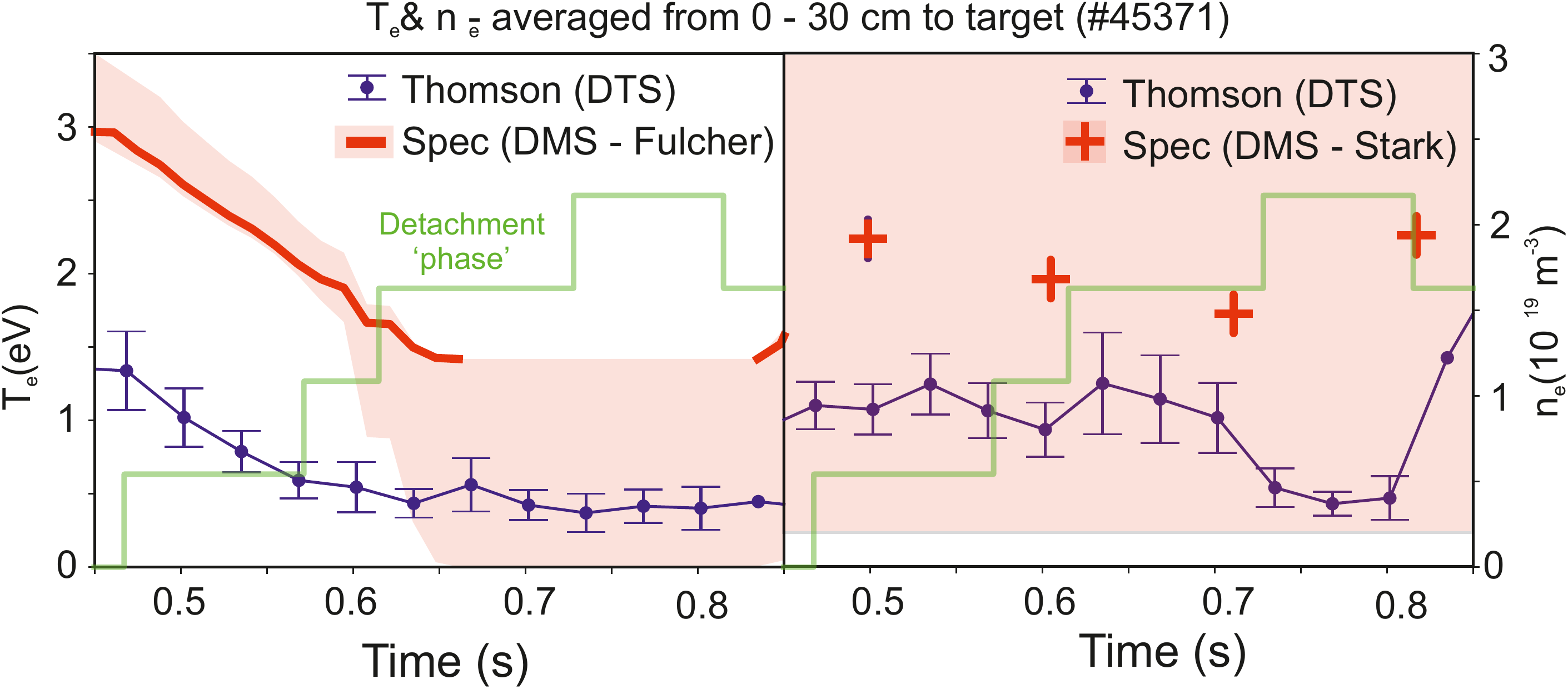}
    \caption{Comparison between the spectroscopically (spec - DMS) inferred densities (Stark) and temperatures (Fulcher brightness) and those obtained from divertor Thomson scattering (DTS) as function of time for \# 45371. To obtain this data, the DTS and spectroscopic inferences have been averaged along the DTS coverage - which spans 0-25 cm from the target. The shaded red region indicates the uncertainty in the spectroscopically derived quantities. As a reference, the $T_e$ measurement limit of the DTS is provided by a grey line and the detachment 'phase' is provided in green (copied from figure \ref{fig:DischargeOverview}).}
    \label{fig:neTeDMSDTS}
\end{figure}

In figure \ref{fig:neTeDMSDTS} we compare the electron density and temperature estimates obtained spectroscopically (from Stark broadening and the Fulcher brightness analysis) and those obtained from divertor Thomson scattering \cite{Clark2021}. Those estimates are obtained by averaging the data close to the target (0-25 cm) and are shown as function of time throughout the discharge, together with the detachment phase \footnote{The Stark inferred electron densities have large uncertainties ($n_e = [0.2 - 3] \times 10^{19} m^ {-3}$ due to the 1) usage of a medium-resolution grating (such that the $n=5,6$ Balmer lines can be covered by the DMS); 2) usage of a relatively low medium-n Balmer line for the Stark broadening ($n=5$), usage of the $n=7$ transition was not possible due to impurity emission contamination; 3) the low electron density conditions resulting in narrow Stark widths compared to the Doppler and instrumental widths (0.01 nm at $n_e = 10^{19} m^{-3}$ vs 0.02 nm (at $T = $ 1 eV) and 0.09 nm respectively).}. The divertor Thomson scattering (DTS) results are local measurements, whereas the BaSPMI $n_e$, $T_e$ inferences are characteristic emission weighted quantities along the DMS lines of sight. The local DTS measurements are sensitive to the exact position of the separatrix with respect to the DTS measurement points to the divertor leg which is slightly misaligned (see figure \ref{fig:GeomLoS}), whereas the DMS inferences are not affected by this as it is a line integral. The inferred electron temperature from spectroscopy is characteristic of the hot temperature regime along the line of sight (as this is the location, in these plasma temperature regimes, where Fulcher emission occurs). The DTS measurements are frequently not well aligned with the separatrix and therefore we expect that the DTS measurements provide lower values for $n_e, T_e$ than the DMS. Higher accuracy $n_e$ and $T_e$ data can be obtained from the high-n Balmer line setup (see section \ref{ch:strong_EIR}), which can be applied from detachment phase III onwards, however in those low temperature conditions, the temperature decays below the DTS threshold ($T_e \leq 0.5$ eV). Once detachment phase 2-3 are obtained, the Fulcher emission region has detached from the target and as such, the Fulcher emission analysis only provides an upper constraint of the possible temperature of 1.3 eV.

A qualitative agreement between the trends of the spectroscopically inferred densities/temperatures and those obtained by DTS is shown. As expected, the DTS measurements provide lower temperatures and densities than the spectroscopic measurements – although the density estimates are within the (very wide) uncertainties of the spectroscopic Stark estimates, which is a systematic uncertainty. The temperature is observed to drop during the discharge with a minor drop in density (detachment phase IV) for both cases. During the gas cut phase, the density is shown to increase for both cases. Both the DTS and the Fulcher inferred temperature estimates indicate low temperatures throughout the discharge ($<3$ eV). The high-n Balmer line fitting provides density estimates with much lower uncertainty (figure \ref{fig:DenFrontMovement}) and has a temperature sensitivity below 1.4 eV. This indicates electron densities more in line with the DTS (see figure \ref{fig:DenFrontMovement}) as well as a presence of $\ll 0.5$ eV temperatures, which is consistent with the DTS reaching its measurement limit of 0.5 eV \cite{Clark2021}.
 
\section{Discussion}
\label{ch:discussion}


\subsection{Understanding the impact of plasma-molecule interactions on hydrogen emission on MAST-U}
\label{ch:importance_plasmamol}

In section \ref{ch:ionisation_movement} (phase I detachment) it was shown that plasma-molecular interactions can have a strong impact on the hydrogen Balmer line emission and its interpretation. This is investigated in more detail by performing simplified emission modelling to investigate chordally integrated Balmer line emission and reaction rate trends as function of the electron temperature. The model uses scalings for the interaction lengths along the line of sight $\Delta L$ times the molecular ($n_{D_2}$) and hydrogen atom ($n_o$) densities relative to the electron densities as function of temperature (e.g. $\Delta L n_{D_2}/n_e$ and $\Delta L n_o/n_e$ respectively), obtained in \ref{ch:emissmodel} from SOLPS-ITER modelling of MAST-U \cite{Myatra} and TCV \cite{Fil2017}. Those molecular and hydrogen atom densities are then combined with a model that post-processes / extrapolates \cite{Verhaegh2021} the hydrogen molecular densities to density estimates for molecular ions using $n_e, T_e$ together with hydrogen rates \cite{Sawada1995,AMJUEL} for the creation and destruction of molecular ions. Coefficients from \cite{Reiter2018} are used to convert those hydrogenic coefficients into coefficients for deuterium (lowering the $D_2^+$ and $D^-$ content by 5 \% and 30 \% respectively with respect to $H_2^+$ and $H^-$). This data is then combined with photon emission coefficients to generate the Balmer line brightnesses and rate data to generate the various chordally integrated rates (ionisation, EIR, etc.). 

Although the studied discharges are Ohmically heated, the used non-interpretive, MAST-U simulations from \cite{Myatra} utilise an input power of 2.5 MW (e.g. 2 MW of beam heating), whilst the investigated MAST-U Super-X scenarios are Ohmic ($\sim 0.6$ MW of Ohmic heating). The TCV simulations are interpretive (Ohmic) simulations. Experimental reference values $n_e = 10^{19} m^{-3}$ were assumed for MAST-U \setcounter{footnote}{0} \footnote{$n_e = 3 \cdot 10^{20} m^{-3}$ would be characteristic for the performed simulations}, whereas $n_e = 7 \cdot 10^{19} m^{-3}$ was taken as a characteristic value for the TCV divertor \cite{Verhaegh2017,Verhaegh2018,Verhaegh2019}. The results, shown in figure \ref{fig:BalmerModel}, are now described:  

\begin{figure}
    \centering
    \includegraphics[width=0.88\linewidth]{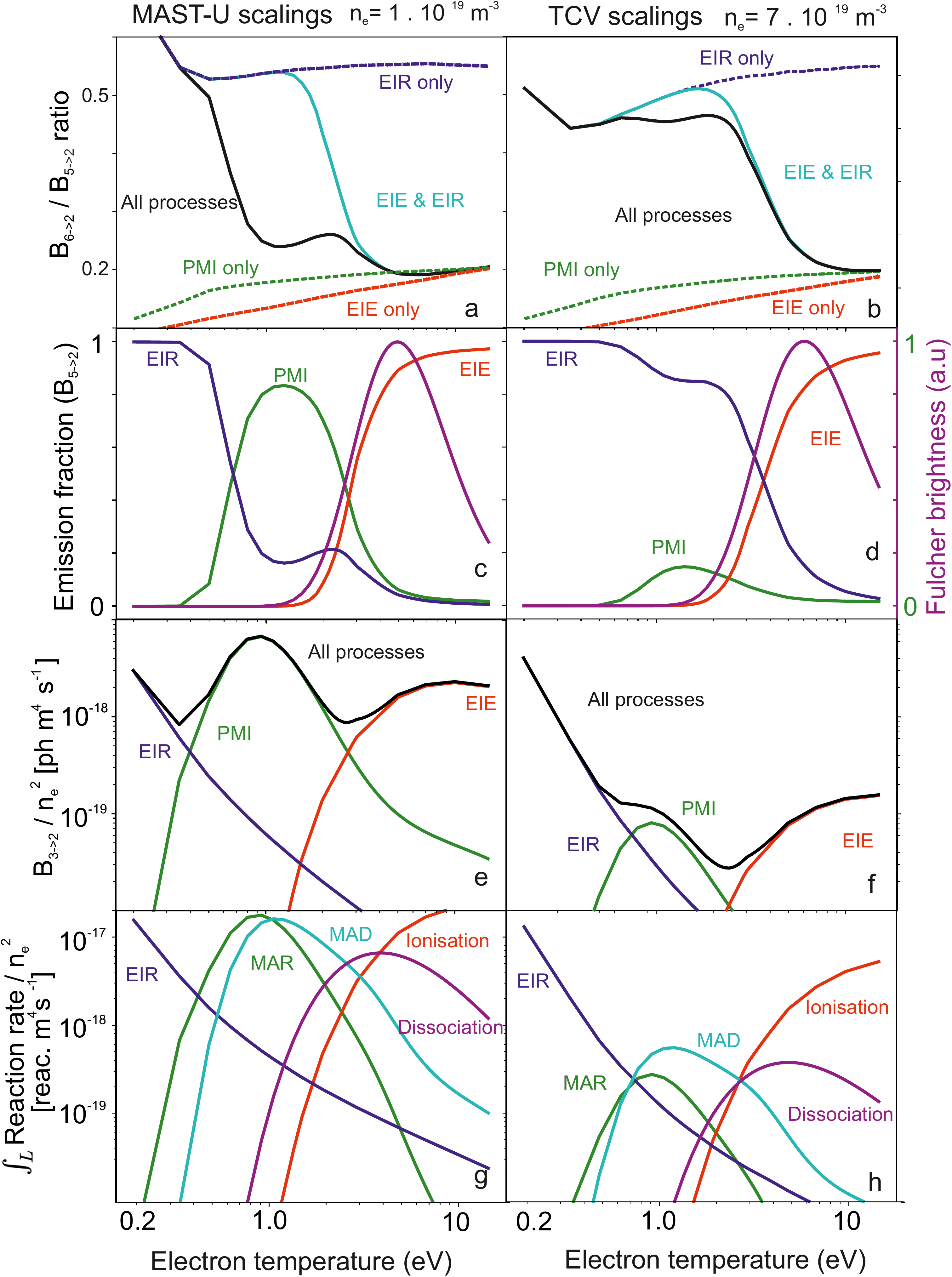}
    \caption{{\scriptsize Simplified Balmer emission modelling at two different electron densities using MAST-U and TCV molecular density scalings, respectively, derived from SOLPS-ITER modelling (\ref{ch:emissmodel}). Note that these scalings for TCV have been derived in an open divertor geometry (before the installation of baffles). a, b) The $n=6 / n=5$ Balmer line ratio is shown as function of $T_e$ when only atomic effects are included (cyan) and when all processes are included (black). Blue, green, red dotted lines show the expected EIR, PMI and EIE only trends. c, d) The $n=5$ Balmer line emission fractions are shown as function of $T_e$ for EIR, PMI and EIE as well as the normalised expected Fulcher emission brightness. e, f) The $D\alpha$ ($n=3$) brightnesses, divided by $n_e^2$, of EIR, PMI and EIE as well as the total are shown as function of $T_e$. g, h) The chordally integrated reaction rate of ionisation, electron-ion recombination (EIR), molecular activated dissociation/dissociation (MAR/MAD) and electron-impact dissociation \setcounter{footnote}{0} \footnote{if the unit of reaction rate $R$ is $reac/m^3 s$, then the unit of the chordally integrated reaction rate is $[ \int_L R dL] = reac./m^2 s$) \cite{Verhaegh2019a,Verhaegh2021}} are divided by the characteristic electron density squared and shown as function of $T_e$.}}
    \label{fig:BalmerModel}
\end{figure}

\begin{itemize}
    \item Figure \ref{fig:BalmerModel} a,b shows PMI leads to line ratios that are very similar to EIE, but lower than EIR. PMI will thus reduce the Balmer line ratio if both EIR and PMI are important. With the MAST-U scalings and the $n_e = 10^{19} m^{-3}$ characteristic electron density for current Ohmic MAST-U operation, PMI reduces the temperature at which the line ratio transitions from a low value (EIE \& PMI) to a high value (EIR) (black curve compared to the cyan curve, figure \ref{fig:BalmerModel} a). This shows that analysing Balmer line emission measurements in the current MAST-U conditions, but neglecting PMI, may lead to gross misinterpretations (up to orders of magnitude) of the ionisation source when $T_e = [0.4 - 7]$ eV. 
    
    \item The Fulcher emission brightness is well correlated with the EIE emission, the $D_2$ electron-impact dissociation rate and the ionisation source. This reinforces its use to separate EIE and PMI related emission (further discussed in section \ref{ch:fulcher_ionisation}). Additionally, comparing figures \ref{fig:BalmerModel} a and c shows that the MAST-U observed separation (in $T_e$ space) between the Fulcher emission and the Balmer line ratio increase regions are expected when including molecular effects. 
    
    \item At temperatures below 1 eV, the PMI related emission (and thus the MAR/MAD rate) is reduced as the likelihood of creating $D_2^+$ from $D_2$ through molecular charge exchange reduces. Together with EIE (and thus ionisation) and EIR, this leads to three different $D\alpha$ emission regions or 'fronts' at constant $n_e$ (figure \ref{fig:BalmerModel} e). This explains why four different detachment phases can be identified based on monitoring the $D\alpha$ emission on MAST-U. 
    
    \item Throughout this paper, we have analysed the EIE, PMI and EIR $D\alpha$ emission quantitatively and assumed that these are representative of the ionisation, MAR/MAD and EIR regions, respectively. Comparing figure \ref{fig:BalmerModel} g, h against e, f provides more quantitative evidence for this assumption: a strong correlation is observed between the: 1) $D\alpha$ EIE emission and ionisation rate; 2) $D\alpha$ EIR emission and EIR rate; 3) $D\alpha$ PMI emission and MAR and MAD rate; 4) the Fulcher band emission and the electron-impact dissociation rate. \footnote{MAI is found to be negligible and exhibits a similar trend as the electron-impact dissociation rate.}
    
    \item MAD occurs at slightly higher temperatures than MAR, but at a significantly smaller temperature than electron-impact dissociation. MAD \footnote{Figure \ref{fig:BalmerModel} shows the total number of MAD reactions, which is significantly smaller than the number of neutral atoms created from plasma-molecule interactions, which is 1-2 (2-3) per reaction for MAD (MAR).} may thus be important in plasmas where the ionisation region is detached from the target as, here, it may be a dominant source of neutrals. This can be identified from a spatial mismatch between the Fulcher (electron-impact dissociation) and Balmer PMI (MAR/MAD) emission regions.
\end{itemize}

The electron density may explain some, but not all of the difference between the MAST-U and TCV observations in \cite{Verhaegh2017,Verhaegh2019,Verhaegh2021,Verhaegh2021b,Perek2021}. Performing simplified emission modelling for MAST-U (figure \ref{fig:BalmerModel} a,c,e,g) at the same density as TCV (figure \ref{fig:BalmerModel} b,d,f,h) leads to a result that is in between the shown MAST-U and TCV cases (not shown). Although it removes the $T_e$ shift of the onset of the line ratio increase, induced by PMI, the 6/5 Balmer line ratio plateaus between the EIR and PMI/EIE dominant levels after its initial increase ($\sim 0.3$) and only increases to EIR dominant levels at $T_e<0.8$ eV. Neglecting PMI would have a smaller impact on the spectroscopic inferences in such conditions, although the ionisation source can still be strongly overestimated. Additionally, the detailed analysis in \cite{Verhaegh2021} shows strong PMI contributions to a synthetic DMS signal after post-processing - on which BaSPMI was tested. Higher electron densities are more relevant for future beam heated MAST-U operation, where the higher input power likely increases the detachment onset point and thus the upstream and divertor electron densities at which detachment occurs. 

There are two other differences between the MAST-U and TCV simulations \& observations. First, the $D_2/D$ ratio is $\sim 2.5$ times higher on MAST-U than TCV according to the scalings derived from the SOLPS-ITER simulations (\ref{ch:emissmodel}). Therefore, PMI is expected to dominate over EIE at higher temperatures on MAST-U than on TCV - even at the same electron density. Secondly, TCV conditions with $T_e\ll 1$ eV have thus far not been diagnosed.

In conclusion, including PMI is important for the modelled TCV conditions to obtain MAR estimates as well as well as a better estimation of the ionisation source during deep detached conditions \cite{Verhaegh2021,Verhaegh2021a,Verhaegh2021b}. In contrast, including PMI in any MAST-U divertor analysis is essential even for basic ionisation estimates from Balmer emission spectroscopy.

To investigate what happens when higher, reactor-relevant, electron densities are used, the simplified model has been applied assuming $n_e = 10^{21} m^{-3}$s. This shows that, although the EIR/MAR ratio increases as expected, MAR still dominates over EIR and ionisation between 0.7 and 2 eV when the MAST-U scalings are used. For both the MAST-U and TCV scalings, the neutral atom source from MAR and MAD is dominant between 0.5 to 3.5 eV and is $> 2$ (TCV) $\& > 4$ (MAST-U) times larger than the neutral atom source from electron-impact dissociation. For the TCV (non-baffled) scalings, EIR is dominant over MAR at $n_e = 10^{21} m^{-3}$. This illustrates that having an elevated molecular density in the divertor chamber is important for the significance of MAR \& MAD; which is achieved on MAST-U through its divertor baffle. 

Using the MAST-U scaling at $n_e = 10^{21} m^{-3}$ the PMI emission contribution to the $5\rightarrow2$ transition exceeds that of electron-impact excitation for $T_e < 2.5$ eV, implying that it can significantly impact ionisation source estimates below 3 eV. However, the total PMI emission contribution is minor (up to 10\%) and hence the ionisation source will likely not be overestimated by orders of magnitude if PMI emission contributions are neglected in a hydrogen emission analysis.

Whether these predictions extend to high power reactor relevant regimes is unknown at this point and depends on how the molecular density and neutral atom density, relative to the electron density, scalings as function of $T_e$ change in higher power regimes - which could be altered due to the shorter mean free paths.


\subsubsection{Caveats to the simplified emission modelling}

An important caveat regarding this analysis is that the divertor simulations for TCV used in this analysis did not have baffles installed. At this time, TCV has the option to install divertor baffles \cite{Reimerdes2021}. That should increase the higher neutral molecule and atom densities and may bring the TCV trends of figure \ref{fig:BalmerModel} more in line with those shown for MAST-U. This prediction requires further investigation.

A second caveat is that the derived scalings from SOLPS-ITER simulations (for molecular densities, atomic densities and emission pathlengths) are based on simulations that utilise the default set of reactions. Previous research has shown that such a reaction set for hydrogen, where the molecular charge exchange rate is mass rescaled to deuterium, results in a strongly underestimated $D_2^+/D_2$ ratio in detachment relevant regimes \cite{Verhaegh2021b}; effectively making the MAR, MAD reactions, as well as the plasma-molecular interaction contribution to the Balmer line emission negligible for TCV. As a post-processing approach is applied to convert the molecular density scalings to molecular ion densities in the simplified emission model, the strongly underestimated $D_2^+/D_2$ ratio in SOLPS-ITER does not impact our results directly. They can, however, affect our results indirectly as including a more correct molecular charge exchange rate self-consistently can result in different scalings for the molecular / neutral atom densities \cite{Williams2022}. Preliminary analysis of these updated simulations shows that they would not alter the molecular density scaling as function of $T_e$ significantly, but they would increase the neutral density by up to a factor 2 at lower temperatures ($T_e < 2.5$ eV) due to the increased level of molecular dissociation. At such low temperatures, however, the impact of the neutral atom density on the hydrogen emission and the ionisation source is negligible. 

A third caveat of the simplified emission model is that the post-processing of converting molecular densities into molecular ion densities is heavily dependent on the distribution of vibrational states as well as the vibrationally resolved molecular charge exchange rates, for which rates from AMJUEL / Eirene / Sawada are used \cite{Sawada1995, AMJUEL, Reiter2008}, see the discussion in section \ref{ch:relevance}. 

\subsection{Usage of the Fulcher emission region to identify movements in the ionisation front}
\label{ch:fulcher_ionisation}

Section \ref{ch:importance_plasmamol} described a correlation between the temperatures, where the Fulcher band emits, and where electron-impact excitation and atomic ionisation occurs. This provides potential to use the Fulcher band brightness along the divertor leg as a diagnostic for the temperature regime and as an identifier for the detachment front edge.
  
Assuming a plasma with constant electron density $n_e$, molecular density $n_{H_2}$ and $T_e$, the Fulcher band brightness monitored by a line of sight intersecting that plasma with intersection length $\Delta L$ can be modelled as: $B_{Fulcher} = \Delta L n_e n_{H_2} PEC (n_e, T_e)$. Combining this Fulcher brightness model with the SOLPS-ITER scalings for $f_{\Delta L n_{D_2} / n_e} \approx \frac{\Delta L n_{D_2}}{n_e}$, which is a parametrised function of $T_e$ (see appendix \ref{ch:emissmodel} for further explanation), the Fulcher band brightness can be modelled as $B_{Fulcher} \approx f_{\Delta L n_{D_2} (Te) / n_e} (T_e) n_e^2 PEC_{Fulcher} (n_e, T_e)$, where $PEC_{Fulcher} (n_e, T_e)$ is a photon emission coefficient for the total Fulcher band emission. Plotting the Fulcher band intensity as function of $T_e$ for a fixed $n_e$ leads to a peaked profile with a trailing (low $T_e$) edge (as shown in figure \ref{fig:BalmerModel} c, d). We find that the normalised shape of the trailing edge of these profiles is only weakly sensitive to the uncertainty of the $f_{\Delta L n_{D_2} / n_e}$ parameter. Therefore, since $T_e$ varies along the divertor leg, it may be possible to utilise the Fulcher band intensity along the divertor leg (e.g. the spatial $B_{Fulcher}$ profile) as a rough temperature constraint under the assumption that either $n_e$ does not change along the divertor leg or by using a measured $n_e$ (e.g. Stark broadening) and observing the spatial $B_{Fulcher}/n_e^2$ profile.  

\begin{figure}
    \centering
    \includegraphics[width=0.8\linewidth]{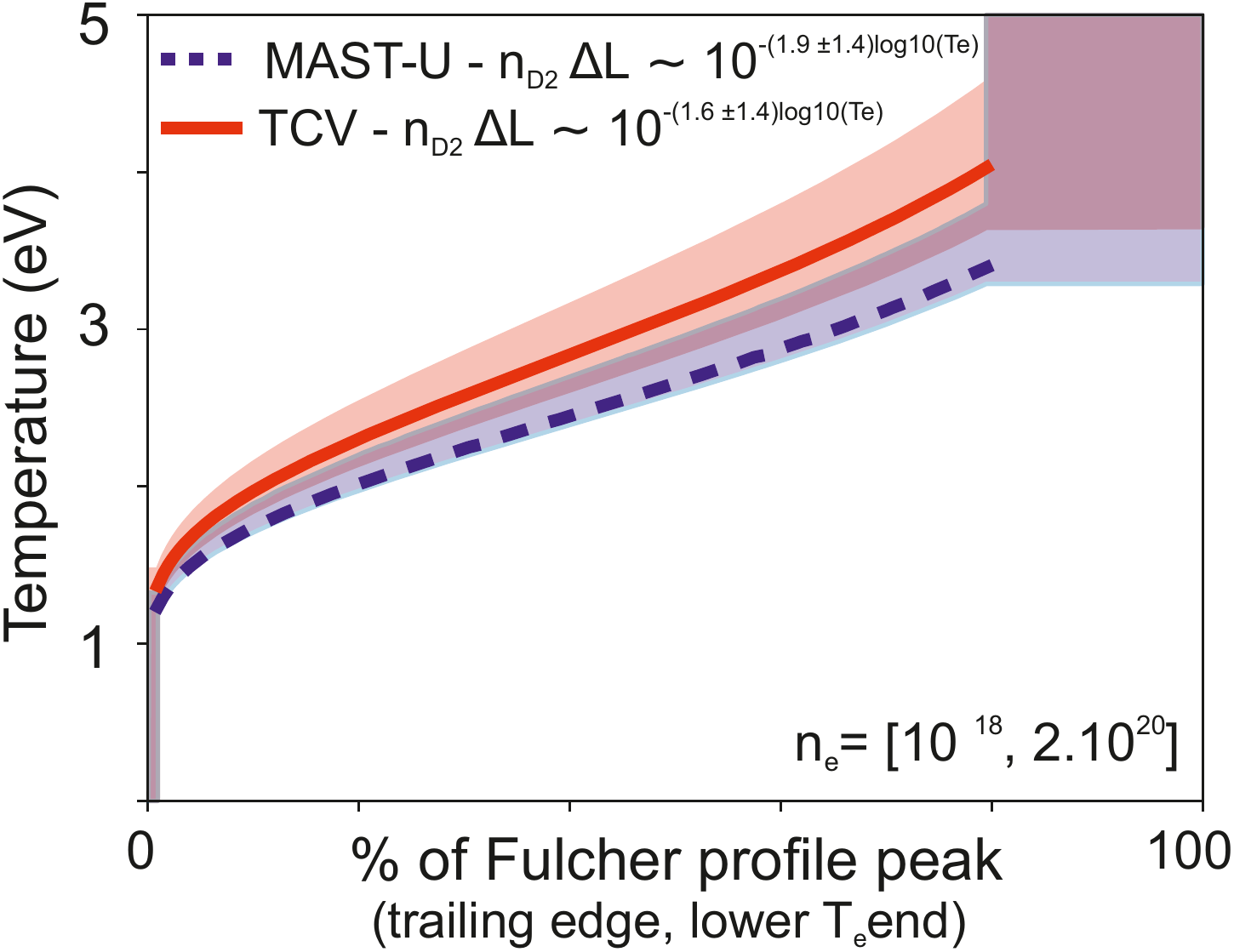}
    \caption{Monte Carlo uncertainty propagation analysis linking the trailing edge of the Fulcher emission profile (between 1 and 80 \% of its peak intensity) to a possible temperature range (68 \% confidence). Below/above this range, it can provide a maximum / minimum temperature constraint, respectively.}
    \label{fig:FulcherTe}
\end{figure}

We investigate this more quantitatively by employing Monte Carlo uncertainty propagation to probe a large uncertainty in the derived $f_{\Delta L n_{D_2} / n_e}$ by multiplying it with $10^{1.4 (R-1) \log{} 10 (T_e)}$ where R is uniformly randomly sampled between 0 and 2.  This results in many $B_{Fulcher} (T_e)$ profiles and for each of the cold-end of these profiles, the temperature corresponding to 1-80 \% of the peak of the profile, to which an additional uncertainty of $\pm 0.4$ eV is added, is stored. These samples can be used as a temperature constraint for the 1.3 - 3.4 eV range (MAST-U). Below / above this range, it can instead provide a maximum / minimum temperature constraint, respectively. The narrow uncertainties obtained in figure \ref{fig:FulcherTe} and the similarity between the curves from MAST-U and TCV suggests it may be possible to infer a rough estimate of the electron temperature using the Fulcher band emission brightness profile along the divertor leg. \footnote{One caveat to this analysis is that it is sensitive to the Fulcher band brightness emission coefficient, here obtained from AMJUEL/Sawada \cite{AMJUEL,Sawada1995}, which requires further investigation in the future. } 

The temperature sensitivity of the Fulcher emission is similar to the temperature sensitivity of the ionisation source and both are correlated (\ref{ch:fulchemiss_ioni}), which may imply that the Fulcher emission region could become a proxy for the ionisation region. If so, it could be used as a real-time observer in conjunction with filtered camera imaging \cite{Ravensbergen2020,Ravensbergen2021,Perek2019submitted} to control the ionisation region in real-time, which would be a major step in detachment control.



\subsection{The generality of the four detachment phases on MAST-U}
\label{ch:generality}

The presented results are a characteristic representation of the results obtained in MAST-U with a Super-X divertor during the first MAST-U campaign, which did not feature beam heated Super-X discharges. 

The generality of the discovery of the four phases of detachment is unknown at this stage. Future higher power operation in MAST-U is expected to increase $P_{SOL}$ and thus the upstream density at which plasma detachment takes place, greatly raising the divertor densities and thus the importance of EIR. Due to this, detachment phases II (e.g. the detachment of the peak in PMI related emission from the target) and III (e.g. the appearance of EIR) could overlap or even swap (e.g. the appearance of EIR occurs before the PMI emission detaches from the target).

Although an in-depth comparison of the conventional divertor and the Super-X divertor is outside of the scope of this work and is ongoing, we have observed that only the onset of PMI emission reaches the separatrix (phase I detachment) before the peak core density is reached and signs of MARFEs occur, although signs of EIR emission and PMI emission region movement can occur in the far-SOL (from the strike point on 'tile 2' up to 'tile 5' see figure \ref{fig:GeomLoS}). Therefore, detachment phases II-IV have not yet occurred at the \emph{separatrix} in the conventional divertor in the scanned density range (up to $n_e/n_{GW} = 0.55$).

Only a limited comparison between MAST-U and other reactors, in terms of a detailed investigation of the plasma chemistry during detachment, is possible as the availability of such an analysis is limited to TCV (prior to the baffle installation). Comparing our results to TCV (pre-baffles), we find that the detachment of the ionisation front from the target upstream at the ion target flux roll-over occur in both devices. Contrastingly, we find that EIR emission appears before the detachment of the PMI emission region, which is not observed \cite{Verhaegh2019}. These differences are related to 1) the higher divertor electron densities; 2) the lower levels of divertor molecular density; 3) the electron temperatures do not drop as low on TCV (before baffles) (see section \ref{ch:emissmodel}). Performing more detailed analysis of plasma-atom/molecular interactions in other devices, such as JET \cite{Karhunen2022,Karhunen2022a}, are ongoing and preliminary results also identify the importance of PMI emission.

To conclude, we do not think that the observation of the various processes, categorised by the four different phases of detachment, is a MAST-U Super-X divertor specific observation. The observed processes can occur more generally. It is, however, the combination of 1) low divertor densities ($n_e = 10^{19} m^{_3}$); 2) the low temperatures achieved in the MAST-U divertor ($T_e \leq 0.2$ eV) that explains the sequence and evolution of the four phases of detachment that is distinctively different from current conventional divertor operation on MAST-U that cannot reach as deep levels of detachment before a MARFE-brought on disruption is triggered.

\subsection{Implications of our findings for detachment 'front' location studies and real-time detachment control}
\label{ch:relevanceFrontControl}

Detachment is a combination of processes that occur in the divertor volume. One method of investigating these processes volumetrically is to ascribe a spatial 'extent' to detachment and define a detachment front location $L_p$, which can be tracked as detachment evolves \cite{Theiler2017,Lipschultz2016,Krasheninnikov2017,Hutchinson1994}. Real-time detachment control can be facilitated by tracking $L_p$ and acting upon the fuelling/seeding in real-time feedback to move $L_p$ towards a required value \cite{Ravensbergen2020,Ravensbergen2021}. Analytic models provide predictions for $L_p$ that can be compared against experiments, as function of magnetic divertor geometry in steady-state, in terms of a single or set of multiple 'physical' control parameter(s), such as upstream electron density, heat flux and impurity fraction \cite{Lipschultz2016}. 

The detachment front, however, is not well defined \cite{Lipschultz2016} and investigations on how it should be defined and diagnosed are ongoing. Examples of 'fronts' used for tracking detachment include: 1) the $CIII$ (465 nm) front location \cite{Theiler2017,Ravensbergen2020}, which physically represents a broad possible electron temperature region (4-12 eV - depending on carbon impurity transport). 2) The total radiation front location \cite{Nakazawa2000, Umansky2017}, that appears to be well correlated with the $CIII$ front on TCV as the radiation in (non-seeded) scenarios is dominated by carbon impurities \cite{Theiler2017}. 3) The $D\alpha$ emission front, which is assumed to be a proxy for the ionisation front \cite{Donne2007,Potzel2012}, which cannot be substantiated  without additional information. Although these three examples are, experimentally, straightforwardly to track, they are not necessarily directly correlated to detachment \cite{Krasheninnikov2017,Lipschultz2016}.

Our analysis on MAST-U suggests the presence of a wide operational regime for detachment in the Super-X configuration, exhibiting four phases. Each phase starts with the formation and movement of a "front" of a certain process from the target upstream. All these fronts move from the target at higher densities (and lower temperatures) than the the $CIII$ (465 nm) front. 

The detachment process seems to start with the movement of the ionisation region of the target (phase I), in agreement with published TCV studies \cite{Verhaegh2019,Verhaegh2021a}. The difficulty of using the ionisation region to track the detachment front is, however, in diagnosing the ionisation region. ITER plans to use the $D\alpha$ emission front (or, if unsuccessful, hydrogen atomic emission spectroscopy) to determine the ionisation front \cite{Donne2007}. Our work shows: 1) one cannot assume $D\alpha$ is dominated by EIE without further information and; 2) if PMI is sufficiently strong, the ionisation region cannot be inferred from hydrogen Balmer emission spectroscopy alone as EIE and PMI cannot be distinguished. This is supported by results from TCV \cite{Verhaegh2021,Verhaegh2021a,Verhaegh2021b}; JET \cite{Karhunen2022, Lomanowski2020} and these MAST-U results, where it was shown that molecules contribute strongly to the $D\alpha$ emission. Section \ref{ch:fulcher_ionisation} showed that tracking the $D_2$ Fulcher band emission may, instead, be a suitable proxy for the ionisation region (possibly in combination with atomic hydrogen spectroscopy), meritorious of further study. 

Additional processes occur during deep detachment, beyond the ionisation front movement (phase I detachment), which can also be tracked: 1) the MAR front, which can be tracked using the $D\alpha$ emission on MAST-U (phase II detachment); 2) the EIR emission "high temperature"-end, which can be monitored by tracking any increase in Balmer line ratio (phase III detachment); 3) the electron density bulk, i.e. tracking the "cold"-end of the high-n ($n\geq7$) Balmer line emission region (phase IV detachment). Each front corresponds to a different process and temperature regime and thus changing the tracked front implies changing the tracked physics process. 

Our work has shown that a single detachment front identifier in the divertor chamber is insufficient to track the entire detached operational range of the MAST-U Super-X divertor. For example, the ionisation front leaves the divertor chamber before the divertor reaches its deepest detached phase. Therefore, the detachment window cannot be measured on MAST-U using diagnostics that operate (solely) in the MAST-U divertor chamber (DMS, MWI) as only 60 \% of the distance from target to x-point is monitored. Instead, tracking the entire detached operational range of the MAST-U Super-X divertor requires multiple detachment front identifiers. That will, thus, change the tracked physics process. Adding additional diagnostics with X-point views to MAST-U, such as a second multi-wavelength imaging system as well as operating at higher elongation, would ameliorate this issue. 


\subsection{Implications, relevance and accuracy of our findings}
\label{ch:relevance}

Our results indicate that plasma-molecular interactions have a particularly strong impact on the MAST-U hydrogen emission. One concern is that such interactions are not properly included in SOLPS-ITER simulations: interactions with molecular ions in a deuterium plasma are essentially negligible for the default SOLPS-ITER reaction setup used \cite{Verhaegh2021b}, which has been attributed to the applied isotope re-scaling of the effective molecular charge exchange rate ($H_2 + H^+ \rightarrow H_2^+ + H$). Such interactions not only contribute as an ion and energy sink, but are also the dominant neutral source between $\sim 0.4$ eV and $\sim 3.5$ eV, based on simplified modelling (section \ref{ch:importance_plasmamol}), where electron impact dissociation is diminished and EIR is not a strong source of neutrals. Therefore, including the corrected reaction rates for molecular processes is also important for modelling the neutral atom content.

Even when including such interactions through modified reaction rates, the vibrational distribution of the molecules, which strongly impacts PMI rates, is not modelled in SOLPS-ITER. Instead, a simplified model is employed to model the vibrational distribution based on the local plasma parameters ($n_e, T_e$) \cite{AMJUEL,Reiter2005}. The vibrational distribution $D_2 (\nu)$ obtained using the reaction set included within Eirene deviates significantly from that of Yacora \cite{Wunderlich2020,Wunderlich2021} as was shown in \cite{Holm2022}. Additional effects, such as interactions between the molecules and the walls, as well as transport of vibrationally excited molecules \cite{Holm2022, Laporta2021}, may lead to further uncertainties in the vibrational distribution. The vibrational distribution must, therefore, be investigated in more detail, both in general and, more specifically, for MAST-U through both simulations and experiments. 

Apart from the treatment of vibrationally excited molecules, the vibrationally resolved molecular charge exchange cross-sections \cite{Ichihara2000} as well as the isotope dependency of the various rates \cite{Reiter2018,Krishnakumar2011} are under debate. For instance, the vibrationally resolved molecular charge exchange cross-sections from \cite{Ichihara2000} result in higher reaction probabilities for vibrationally excited molecules at low temperatures $T<1.5$ eV than those in AMJUEL / Eirene / Sawada as it includes $H_2 + H^+ \rightarrow H_3^+ \rightarrow H_2^+ + H$ which is more dominant at low temperatures. This, for instance, would lead to an enhanced MAR/MAD rate and PMI emission contribution at $T<1$ eV compared to the calculations performed in section \ref{ch:importance_plasmamol} if there is a significant population of vibrationally excited molecules.

The achievement of 'stable' operation \footnote{The evidence for this is: 1) operation at a significantly lower core density than where MARFEs occur; 2) cutting the fuelling results in a movement of the detachment regions back towards the target within 50 ms.} with very low divertor temperatures ($\ll 0.5$ eV) may provide some preliminary anecdotal evidence of the Super-X divertor; although further research is required. Those low temperature regimes also have implications for requirements on atomic/molecular data development. By default, ADAS data is available down to 0.2 eV \cite{OMullane,Summers2006}. This results in additional uncertainty in the atomic/molecular physics coefficients used in our BaSPMI analysis, which employs nearest neighbour extrapolation (section \ref{ch:baspmi}). The appearance of low electron temperatures also has implications for plasma-edge codes such as SOLPS-ITER that employs both ADAS and AMJUEL data. Further investigations of obtaining ADAS data at lower temperatures are underway.

As plasma-edge simulations extrapolate our understanding to reactor-class devices, predictions of the impact of molecular effects for reactors are highly uncertain. Simplified modelling may suggest that MAR/MAD can be important at high density regimes if the molecular density in the divertor can be kept sufficiently high (section \ref{ch:importance_plasmamol}). However, whether MAR and MAD contribute significantly below the ionisation region in reactors with or without divertor baffles warrants further study. Reactor concepts with possibly tightly baffled alternative divertor concepts (ADC) (e.g. SPARC \cite{Kuang2020}, ARC \cite{Wigram2019}, STEP \cite{Wilson2020} and DEMO \cite{Militello2021} (potentially)) are likely more impacted by these plasma-molecular interactions as: 1) the baffled region may increase the molecular density, as indicated in section \ref{ch:importance_plasmamol}; 2) as the divertor leg is longer, there can be a larger volume in which plasma-molecular interactions occur; 3) their operational regime may feature deeper detachment with the ionisation region significantly lifted off the target.

\section{Conclusions}
\label{ch:conclusion} 

Detachment of the MAST-U Super-X divertor can be described by four phases: First, the ionisation front detaches from the target, leading to the formation of a region where the molecular density increases and these molecules become vibrationally excited. The resulting molecular ions ($D_2^+$ and/or $D_2^- \rightarrow D^- + D$) subsequently react with the plasma through Molecular Activated Recombination and Dissociation (MAR/MAD), which generates excited atoms that emit atomic hydrogen line emission. Secondly, the peak in hydrogen line emission from those plasma-molecular interactions detaches from the target, leaving a region where the plasma is too cold to promote molecular ion generation ($T_e \leq \sim 0.7$ eV). Thirdly, strong contributions of electron-ion recombination to the hydrogen line emission start to develop. Finally, the emissive contributions from electron-ion recombination start to detach from the target, which is hypothesised to be related to the displacement of the electron density bulk from the target. High-n Balmer line fitting facilitates higher accuracy density/temperature estimates, showing a clear drop in the electron density near the target during deep detachment, in agreement with this hypothesis. The description of the four phases of detachment was motivated through quantitative analysis and by simulating the expected Balmer line emission behaviour using a simplified emission model. Evidence of strongly sub-eV temperatures is presented during phases III and IV, with preliminary evidence of $T_e < 0.2$ eV. The electron density is modest throughout phases 1-3 ($n_e < 2 \times 10^{19} m^{-3}$), but can descend to levels of $n_e \ll 1 \times 10^{19} m^{_3}$ in phase IV below the density bulk.

We have demonstrated the utility of using $D_2$ Fulcher brightness and shown that it can be used as a temperature indicator and possibly as a proxy for the ionisation region. That information, combined with spectroscopic analysis, facilitated separating the electron-impact excitation emission from emission arising from excited hydrogen atoms after molecular break-up involving molecular ions. As the contribution of plasma-molecule interactions to the hydrogen Balmer line emission is particularly strong for MAST-U, accounting for this in the interpretation of hydrogen emission diagnostics is critical, which has implications for diagnostic analysis as well as for the development of observers for tracking the detachment front. Plasma-molecule interactions that involve MAR and MAD appear to play a strong role in the detached MAST-U Super-X divertor.

\section{Acknowledgements}

This work has been carried out within the framework of the EUROfusion Consortium, funded by the European Union via the Euratom Research and Training Programme (Grant Agreement No 101052200 — EUROfusion), from the RCUK Energy Programme and EPSRC Grants EP/T012250/1 and EP/N023846/1. It has been supported in part by the Swiss National Science Foundation. To obtain further information on the data
and models underlying this paper please contact publicationsmanager@ukaea.uk. Views and opinions expressed are however those of the author(s) only and do not necessarily reflect those of the European Union or the European Commission. Neither the European Union nor the European Commission can be held responsible for them.

\section{References}

\bibliographystyle{iopart-num}
\bibliography{all_bib.bib}

\providecommand{\newblock}{}
\begin{thebibliography}{10}
\expandafter\ifx\csname url\endcsname\relax
  \def\url#1{{\tt #1}}\fi
\expandafter\ifx\csname urlprefix\endcsname\relax\def\urlprefix{URL }\fi
\providecommand{\eprint}[2][]{\url{#2}}

\bibitem{Loarte2007}
Loarte A, Lipschultz B, Kukushkin A, Matthews G, Stangeby P, Asakura N,
  Counsell G, Federici G, Kallenbach A, Krieger K {\em et~al.\/} 2007 {\em
  Nuclear Fusion\/} {\bf 47} S203--S263

\bibitem{Wenninger2014}
Wenninger R, Bernert M, Eich T, Fable E, Federici G, Kallenbach A, Loarte A,
  Lowry C, McDonald D, Neu R {\em et~al.\/} 2014 {\em Nuclear Fusion\/} {\bf
  54} 114003

\bibitem{Verhaegh2021b}
Verhaegh K, Lipschultz B, Harrison J, Duval B, Fil A, Wensing M, Bowman C,
  Gahle D, Kukushkin A, Moulton D, Perek A, Pshenov A, Federici F,
  F{\'{e}}vrier O, Myatra O, Smolders A, Theiler C, the TCV~Team and the
  EUROfusion MST1~Team 2021 {\em Nuclear Fusion\/} {\bf 61} 106014
  \urlprefix\url{https://doi.org/10.1088/1741-4326/ac1dc5}

\bibitem{Lipschultz2016}
Lipschultz B, L~Para F and Hutchinson I 2016 {\em Nuclear Fusion\/} {\bf 56}
  056007 ISSN 0029-5515
  \urlprefix\url{http://stacks.iop.org/0029-5515/56/i=5/a=056007}

\bibitem{Moulton2017}
Moulton D, Harrison J, Lipschultz B and Coster D 2017 {\em Plasma Physics and
  Controlled Fusion\/} {\bf 59} 065011 ISSN 0741-3335

\bibitem{Havlickova2015}
Havlickova E, Harrison J, Lipschultz B, Fishpool G, Kirk A, Thornton A,
  Wischmeier M, Elmore S and Allan S 2015 {\em Plasma Physics and Controlled
  Fusion\/} {\bf 57} 115001 ISSN 0741-3335 \urlprefix\url{<Go to
  ISI>://WOS:000374538100002
  http://iopscience.iop.org/article/10.1088/0741-3335/57/11/115001/pdf}

\bibitem{Theiler2017}
Theiler C, Lipschultz B, Harrison J, Labit B, Reimerdes H, Tsui C, Vijvers
  W~A~J, Boedo J~A, Duval B~P, Elmore S, Innocente P, Kruezi U, Lunt T,
  Maurizio R, Nespoli F, Sheikh U, Thornton A~J, van Limpt S~H~M, Verhaegh K,
  Vianello N, Team T and Team E~M 2017 {\em Nuclear Fusion\/} {\bf 57} 072008
  ISSN 0029-5515 \urlprefix\url{<Go to ISI>://WOS:000398746200001
  http://iopscience.iop.org/article/10.1088/1741-4326/aa5fb7/pdf}

\bibitem{Valanju2009}
Valanju P~M, Kotschenreuther M, Mahajan S~M and Canik J 2009 {\em Physics of
  Plasmas (1994-present)\/} {\bf 16} 056110

\bibitem{Militello2021}
Militello F, Aho-Mantila L, Ambrosino R, Body T, Bufferand H, Calabro G,
  Ciraolo G, Coster D, {Di Gironimo} G, Fanelli P {\em et~al.\/} 2021 {\em
  Nuclear Materials and Energy\/} {\bf 26} 100908 ISSN 2352-1791

\bibitem{Morris2018}
Morris W, Harrison J, Kirk A, Lipschultz B, Militello F, Moulton D and Walkden
  N 2018 {\em IEEE Transactions on Plasma Science\/} {\bf 46} 1217--1226

\bibitem{Fishpool2013}
Fishpool G, Canik J, Cunningham G, Harrison J, Katramados I, Kirk A, Kovari M,
  Meyer H and Scannell R 2013 {\em Journal of Nuclear Materials\/} {\bf 438}
  S356--S359 ISSN 0022-3115 proceedings of the 20th International Conference on
  Plasma-Surface Interactions in Controlled Fusion Devices
  \urlprefix\url{https://www.sciencedirect.com/science/article/pii/S0022311513000755}

\bibitem{Harrison2019}
Harrison J, Akers R, Allan S, Allcock J, Allen J, Appel L, Barnes M, Ayed N~B,
  Boeglin W, Bowman C {\em et~al.\/} 2019 {\em Nuclear Fusion\/} {\bf 59}
  112011

\bibitem{Loarte2001}
Loarte A 2001 {\em Plasma Physics and Controlled Fusion\/} {\bf 43} R183--R224
  ISSN 0741-3335 \urlprefix\url{<Go to ISI>://WOS:000169442600001
  http://iopscience.iop.org/article/10.1088/0741-3335/43/6/201/pdf}

\bibitem{Petrie2013}
Petrie T, Canik J, Lasnier C, Leonard A, Mahdavi M, Watkins J, Fenstermacher M,
  Ferron J, Groebner R, Hill D, Hyatt A, Holcomb C, Luce T, Makowski M, Moyer
  R, Osborne T and Stangeby P 2013 {\em Nuclear Fusion\/} {\bf 53} 113024
  \urlprefix\url{https://doi.org/10.1088/0029-5515/53/11/113024}

\bibitem{Fil2019submitted}
Fil A, Lipschultz B, Moulton D, Dudson B~D, F{\'{e}}vrier O, Myatra O, Theiler
  C, Verhaegh K, Wensing M and and 2020 {\em Plasma Physics and Controlled
  Fusion\/} {\bf 62} 035008

\bibitem{Scannell2022}
{R Scannell on behalf of the MAST Upgrade team} 2022 Developing understanding
  of spherical tokamaks with mast upgrade presented at the 48th EPS Conference
  on Plasma Physics Conference, virtual, 2022

\bibitem{Thornton2022}
{A Thornton on behalf of the EuroFusion WPTE and MAST Upgrade and TCV teams}
  2022 Overview of the first results on the performance of the super-x divertor
  on mast-u and comparison with tcv presented at the 25th International
  Conference on Plasma Surface Interactions, virtual, 2022

\bibitem{Kuang2020}
Kuang A~Q, Ballinger S, Brunner D, Canik J, Creely A~J, Gray T, Greenwald M,
  Hughes J~W, Irby J, LaBombard B and et~al 2020 {\em Journal of Plasma
  Physics\/} {\bf 86} 865860505

\bibitem{Wigram2019}
Wigram M, LaBombard B, Umansky M, Kuang A, Golfinopoulos T, Terry J, Brunner D,
  Rensink M, Ridgers C and Whyte D 2019 {\em Nuclear Fusion\/} {\bf 59} 106052
  \urlprefix\url{https://doi.org/10.1088/1741-4326/ab394f}

\bibitem{Wilson2020}
Wilson H, Chapman I, Denton T, Morris W, Patel B, Voss G, Waldon C and the
  STEP~Team 2020 Step—on the pathway to fusion commercialization {\em
  Commercialising Fusion Energy\/} 2053-2563 (IOP Publishing) pp 8--1 to 8--18
  ISBN 978-0-7503-2719-0
  \urlprefix\url{https://dx.doi.org/10.1088/978-0-7503-2719-0ch8}

\bibitem{Verhaegh2021a}
Verhaegh K, Lipschultz B, Harrison J~R, Duval B~P, Bowman C, Fil A, Gahle D~S,
  Moulton D, Myatra O, Perek A, Theiler C and Wensing M 2021 {\em Nuclear
  Materials and Energy\/} {\bf 26} 100922

\bibitem{Verhaegh2021}
Verhaegh K, Lipschultz B, Bowman C, Duval B~P, Fantz U, Fil A, Harrison J~R,
  Moulton D, Myatra O, Wünderlich D, Federici F, Gahle D~S, Perek A, Wensing M
  and and 2021 {\em Plasma Physics and Controlled Fusion\/} {\bf 63} 035018

\bibitem{Verhaegh2019a}
Verhaegh K, Lipschultz B, Duval B, Fil A, Wensing M, Bowman C and Gahle D 2019
  {\em Plasma Phys. Control. Fusion\/} {\bf 61}

\bibitem{Lomanowski2020}
Lomanowski B, Groth M, Coffey I~H, Karhunen J, Maggi C~F, Meigs A, Menmuir S
  and O'Mullane M 2020 {\em Plasma Physics and Controlled Fusion\/} {\bf 62}

\bibitem{Karhunen2022}
Karhunen J, Holm A, Lomanowski B, Solokha V, Aleiferis S, Carvalho P, Groth M,
  Lawson K, Meigs A and Shaw A 2022 {\em Journal of Instrumentation\/} {\bf 17}
  C01032 \urlprefix\url{https://doi.org/10.1088/1748-0221/17/01/c01032}

\bibitem{Karhunen2022a}
Karhunen J, Holm A, Lomanowski B, Solokha V, Aleiferis S, Carvalho P, Groth M,
  Lawson K, Meigs A, Shaw A {\em et~al.\/} 2022 {\em Plasma Physics and
  Controlled Fusion\/} {\bf 64} 075001

\bibitem{Clark2021}
Clark J~G, Bowden M~D and Scannell R 2021 {\em Review of Scientific
  Instruments\/} {\bf 92} 043545

\bibitem{Myatra}
Myatra O 2021 {\em Numerical modelling of detached plasmas in the MAST Upgrade
  super-X divertor\/} Ph.D. thesis University of York
  \urlprefix\url{https://etheses.whiterose.ac.uk/29934/1/OMyatra\_thesis.pdf}

\bibitem{Verhaegh2017}
Verhaegh K, Lipschultz B, Duval B~P, Harrison R, Reimerdes H, Theiler C, Labit
  B, Maurizio R, Marini C, Nespoli F, Sheikh U, Tsui C~K, Vianello N, Vijvers
  W~A~J and Team T~T~~E~M 2017 {\em Nuclear Materials and Energy\/} {\bf 12}
  1112--1117 ISSN 2352-1791

\bibitem{Verhaegh2019}
Verhaegh K, Lipschultz B, Duval B, Février O, Fil A, Theiler C, Wensing M,
  Bowman C, Gahle D, Harrison J {\em et~al.\/} 2019 {\em Nuclear Fusion\/} {\bf
  59}

\bibitem{Reiter2018}
Janev R~K and Reiter D 2018 Isotope effects in molecule assisted recombination
  and dissociation in divertor plasmas J\"{u}lich report - juel 4411
  Forschungszentrum Jülich GmbH Jülich englisch
  \urlprefix\url{https://juser.fz-juelich.de/record/850290/files/J\%C3\%BCl\_4411\_Reiter.pdf?version=1}

\bibitem{Holm2022}
Holm A, Wünderlich D, Groth M and Börner P 2022 {\em Contributions to Plasma
  Physics\/} {\bf n/a} e202100189

\bibitem{AMJUEL}
Reiter D 2000 The data file {AMJUEL}: Additional atomic and molecular data for
  {EIRENE} Tech. rep. Forschungszentrum Jülich GmbH
  \urlprefix\url{http://www.eirene.de/html/amjuel.html}

\bibitem{Reiter2008}
Reiter D {\em et~al.\/} 2008 The eirene code user manual Report
  Forschungszentrum Jülich GmbH
  \urlprefix\url{http://www.eirene.de/manuals/eirene.pdf}

\bibitem{Ichihara2000}
Ichihara A, Iwamoto O and Janev R~K 2000 {\em Journal of Physics B: Atomic,
  Molecular and Optical Physics\/} {\bf 33} 4747--4758

\bibitem{Sawada1995}
Sawada K and Fujimoto T 1995 {\em Journal of applied physics\/} {\bf 78}
  2913--2924

\bibitem{Lomanowski2015}
Lomanowski B~A, Meigs A~G, Sharples R~M, Stamp M, Guillemaut C and Contributors
  J 2015 {\em Nuclear Fusion\/} {\bf 55} 123028 ISSN 0029-5515
  \urlprefix\url{<Go to ISI>://WOS:000366534500030
  http://iopscience.iop.org/article/10.1088/0029-5515/55/12/123028/pdf}

\bibitem{Lipschultz1998}
Lipschultz B, Terry J~L, Boswell C, Hubbard A, LaBombard B and Pappas D~A 1998
  {\em Physical Review Letters\/} {\bf 81} 1007--1010 ISSN 0031-9007
  \urlprefix\url{<Go to ISI>://WOS:000075130400019
  https://journals.aps.org/prl/abstract/10.1103/PhysRevLett.81.1007}

\bibitem{Terry1998}
Terry J~L, Lipschultz B, Pigarov A~Y, Krasheninnikov S~I, LaBombard B, Lumma D,
  Ohkawa H, Pappas D and Umansky M 1998 {\em Physics of Plasmas\/} {\bf 5}
  1759--1766 ISSN 1070-664x

\bibitem{Kunze2009}
Kunze H~J 2009 {\em Introduction to plasma spectroscopy\/} vol~56 (Springer)

\bibitem{OMullane}
O’Mullane M 2013 Adas: Generalised collisonal radiative data for hydrogen
  Tech. rep. ADAS \urlprefix\url{http://www.adas.ac.uk}

\bibitem{Feng2021}
Feng X, Calcines A, Sharples R~M, Lipschultz B, Perek A, Vijvers W~A~J,
  Harrison J~R, Allcock J~S, Andrebe Y, Duval B~P and Mumgaard R~T 2021 {\em
  Review of Scientific Instruments\/} {\bf 92} 063510 (\textit{Preprint}
  \eprint{https://doi.org/10.1063/5.0043533})
  \urlprefix\url{https://doi.org/10.1063/5.0043533}

\bibitem{Perek2019submitted}
Perek A, Vijvers W~A~J, Andrebe Y, Classen I~G~J, Duval B~P, Galperti C,
  Harrison J~R, Linehan B~L, Ravensbergen T, Verhaegh K and de~Baar M~R 2019
  {\em Review of Scientific Instruments\/} {\bf 90} 123514

\bibitem{Perek2021}
Perek A, Linehan B, Wensing M, Verhaegh K, Classen I, Duval B, Février O,
  Reimerdes H, Theiler C, Wijkamp T and {de Baar} M 2021 {\em Nuclear Materials
  and Energy\/} {\bf 26} 100858 ISSN 2352-1791

\bibitem{Perek2022}
Perek A, Wensing M, Verhaegh K, Linehan B, Reimerdes H, Bowman C, van Berkel M,
  Classen I, Duval B, F{\'{e}}vrier O, Koenders J, Ravensbergen T, Theiler C,
  de~Baar M, the EUROfusion MST1~Team and the TCV~Team 2022 {\em Nuclear
  Fusion\/} {\bf 62} 096012
  \urlprefix\url{https://doi.org/10.1088/1741-4326/ac7813}

\bibitem{Fil2017}
Fil A~M~D, Dudson B~D, Lipschultz B, Moulton D, Verhaegh K~H~A, Fevrier O and
  Wensing M 2017 {\em Contributions to plasma physics\/} {\bf 58} ISSN
  0863-1042

\bibitem{Verhaegh2018}
Verhaegh K 2018 {\em Spectroscopic Investigations of detachment on {TCV}\/}
  Thesis University of York
  \urlprefix\url{http://etheses.whiterose.ac.uk/22523/}

\bibitem{Reimerdes2021}
Reimerdes H, Duval B, Elaian H, Fasoli A, F{\'{e}}vrier O, Theiler C, Bagnato
  F, Baquero-Ruiz M, Blanchard P, Brida D {\em et~al.\/} 2021 {\em Nuclear
  Fusion\/} {\bf 61} 024002

\bibitem{Williams2022}
Williams A~C 2022 Investigation of atomic and molecular rates in plasma-edge
  simulations through experiment-simulation comparisons
  \urlprefix\url{https://arxiv.org/abs/2205.12715}

\bibitem{Ravensbergen2020}
Ravensbergen T, van Berkel M, Silburn S~A, Harrison J~R, Perek A, Verhaegh K,
  Vijvers W~A~J, Theiler C, Kirk A and de~Baar M 2020 {\em Nuclear Fusion\/}
  \urlprefix\url{https://doi.org/10.1088\%2F1741-4326\%2Fab8183}

\bibitem{Ravensbergen2021}
Ravensbergen T, van Berkel M, Perek A, Galperti C, Duval B, F{\'e}vrier O, van
  Kampen R, Felici F, Lammers J, Theiler C {\em et~al.\/} 2021 {\em Nature
  communications\/} {\bf 12} 1--9

\bibitem{Krasheninnikov2017}
Krasheninnikov S~I and Kukushkin A~S 2017 {\em Journal of Plasma Physics\/}
  {\bf 83} 155830501 ISSN 0022-3778

\bibitem{Hutchinson1994}
Hutchinson I~H 1994 {\em Nuclear Fusion\/} {\bf 34} 1337--1348 ISSN 0029-5515

\bibitem{Nakazawa2000}
Nakazawa S, Nakajima N, Okamoto M and Ohyabu N 2000 {\em Plasma Physics and
  Controlled Fusion\/} {\bf 42} 401--413
  \urlprefix\url{https://doi.org/10.1088/0741-3335/42/4/303}

\bibitem{Umansky2017}
Umansky M, Rensink M, Rognlien T, LaBombard B, Brunner D, Terry J and Whyte D
  2017 {\em Nuclear Materials and Energy\/} {\bf 12} 918--923 ISSN 2352-1791
  proceedings of the 22nd International Conference on Plasma Surface
  Interactions 2016, 22nd PSI
  \urlprefix\url{https://www.sciencedirect.com/science/article/pii/S2352179116301259}

\bibitem{Donne2007}
Donn{\'{e}} A, Costley A, Barnsley R, Bindslev H, Boivin R, Conway G, Fisher R,
  Giannella R, Hartfuss H, von Hellermann M, Hodgson E, Ingesson L, Itami K,
  Johnson D, Kawano Y, Kondoh T, Krasilnikov A, Kusama Y, Litnovsky A, Lotte P,
  Nielsen P, Nishitani T, Orsitto F, Peterson B, Razdobarin G, Sanchez J, Sasao
  M, Sugie T, Vayakis G, Voitsenya V, Vukolov K, Walker C, Young K and the ITPA
  Topical Group~on Diagnostics 2007 {\em Nuclear Fusion\/} {\bf 47} S337--S384
  \urlprefix\url{https://doi.org/10.1088/0029-5515/47/6/s07}

\bibitem{Potzel2012}
Potzel S 2012 {\em Experimental classification of divertor detachment\/} Thesis
  Technische Universit\"{a}t M\"{u}nchen (TUM)

\bibitem{Reiter2005}
Reiter D, Baelmans M and Börner P 2005 {\em Fusion Science and Technology\/}
  {\bf 47} 172--186 ISSN 1536-1055

\bibitem{Wunderlich2020}
W\"{u}nderlich D, Giacomin M, Ritz R and Fantz U 2020 {\em Journal of
  Quantitative Spectroscopy and Radiative Transfer\/} {\bf 240} 106695 ISSN
  0022-4073

\bibitem{Wunderlich2021}
Wünderlich D, Scarlett L~H, Briefi S, Fantz U, Zammit M~C, Fursa D~V and Bray
  I 2021 {\em Journal of Physics D: Applied Physics\/} {\bf 54} 115201
  \urlprefix\url{https://doi.org/10.1088/1361-6463/abccf2}

\bibitem{Laporta2021}
Laporta V, Agnello R, Fubiani G, Furno I, Hill C, Reiter D and Taccogna F 2021
  {\em Plasma Physics and Controlled Fusion\/}

\bibitem{Krishnakumar2011}
Krishnakumar E, Denifl S, \ifmmode \check{C}\else
  \v{C}\fi{}ade\ifmmode~\check{z}\else \v{z}\fi{} I, Markelj S and Mason N~J
  2011 {\em Phys. Rev. Lett.\/} {\bf 106}(24) 243201

\bibitem{Summers2006}
Summers H~P, Dickson W~J, O'Mullane M~G, Badnell N~R, Whiteford A~D, Brooks
  D~H, Lang J, Loch S~D and Griffin D~C 2006 {\em Plasma Physics and Controlled
  Fusion\/} {\bf 48} 263--293 ISSN 0741-3335 1361-6587

\bibitem{Stangeby2017}
Stangeby P~C and Chaofeng S 2017 {\em Nuclear Fusion\/} {\bf 57} 056007 ISSN
  0029-5515

\bibitem{Wuenderlich2016}
Wünderlich D and Fantz U 2016 {\em Atoms\/} {\bf 4} ISSN 2218-2004

\bibitem{Wensing2019}
Wensing M, Duval B, Fevrier O, Fil A, Galassi D, Havlickova E, Perek A,
  Reimerdes H, Theiler C, Verhaegh K and Wischmeier M 2019 {\em Plasma Phys.
  Control. Fusion\/} {\bf 61}

\bibitem{Lomanowski2019}
Lomanowski B, Carr M, Field A, Groth M, Henderson S, Harrison J, Huber A,
  Jarvinen A, Lawson K, Lowry C {\em et~al.\/} 2019 {\em Nuclear Materials and
  Energy\/} {\bf 20}

\bibitem{Soukhanovskii2022}
Soukhanovskii V, Khrabry A, Scott H~A, Rognlien T, Moulton D and Harrison J~R
  2022 {\em Nuclear Fusion\/}
  \urlprefix\url{http://iopscience.iop.org/article/10.1088/1741-4326/ac6285}

\end{thebibliography}

\appendix

\section{Simplified emission modelling}
\label{ch:emissmodel}

The simplified emission trends shown in figure \ref{fig:BalmerModel} have been modelled using scalings obtained from SOLPS-ITER simulations for the behaviour of the molecular density, neutral atom density and the respective pathlengths as function of $T_e$. Using those scalings with a plasma slab model (such as employed by BaSPMI \cite{Verhaegh2021,Verhaegh2019a,Verhaegh2017}), the hydrogen line brightnesses for electron-impact excitation (EIE), electron-ion recombination (EIR) and plasma-molecule interactions (PMI) can be calculated as indicated in equation \ref{eq:emiss_slab}, where $f_{\Delta L n_D/n_e}$, $f_{\Delta L_{rec}}$ and $f_{\Delta L n_{D_2}/n_e}$ are the scalings used. Those scalings are analogously derived as in \cite{Verhaegh2019a}.

\begin{equation}
\begin{aligned}
    B_{n\rightarrow2}^{EIE} &= n_e^2 (\Delta L n_H / n_e) PEC_{n\rightarrow2}^{EIE} (n_e, T_e) = n_e^2 f_{\Delta L n_D/n_e} (T_e) PEC_{n\rightarrow2}^{EIE} (n_e, T_e) \\
    B_{n\rightarrow2}^{EIR} &= n_e^2 (\Delta L) PEC_{n\rightarrow2}^{EIR} (n_e, T_e) = n_e^2 f_{\Delta L_{rec}} (T_e) PEC_{n\rightarrow2}^{EIR} (n_e, T_e) \\
    B_{n\rightarrow2}^{PMI} &= n_e^2 (\Delta L n_{D_2} / n_e) PEC_{n\rightarrow2}^{PMI} (n_e, T_e) = n_e^2 f_{\Delta L n_{D_2}/n_e} (T_e) PEC_{n\rightarrow2}^{PMI} (n_e, T_e) \label{eq:emiss_slab}
\end{aligned}
\end{equation}

\begin{figure}
    \centering
    \includegraphics[width=\linewidth]{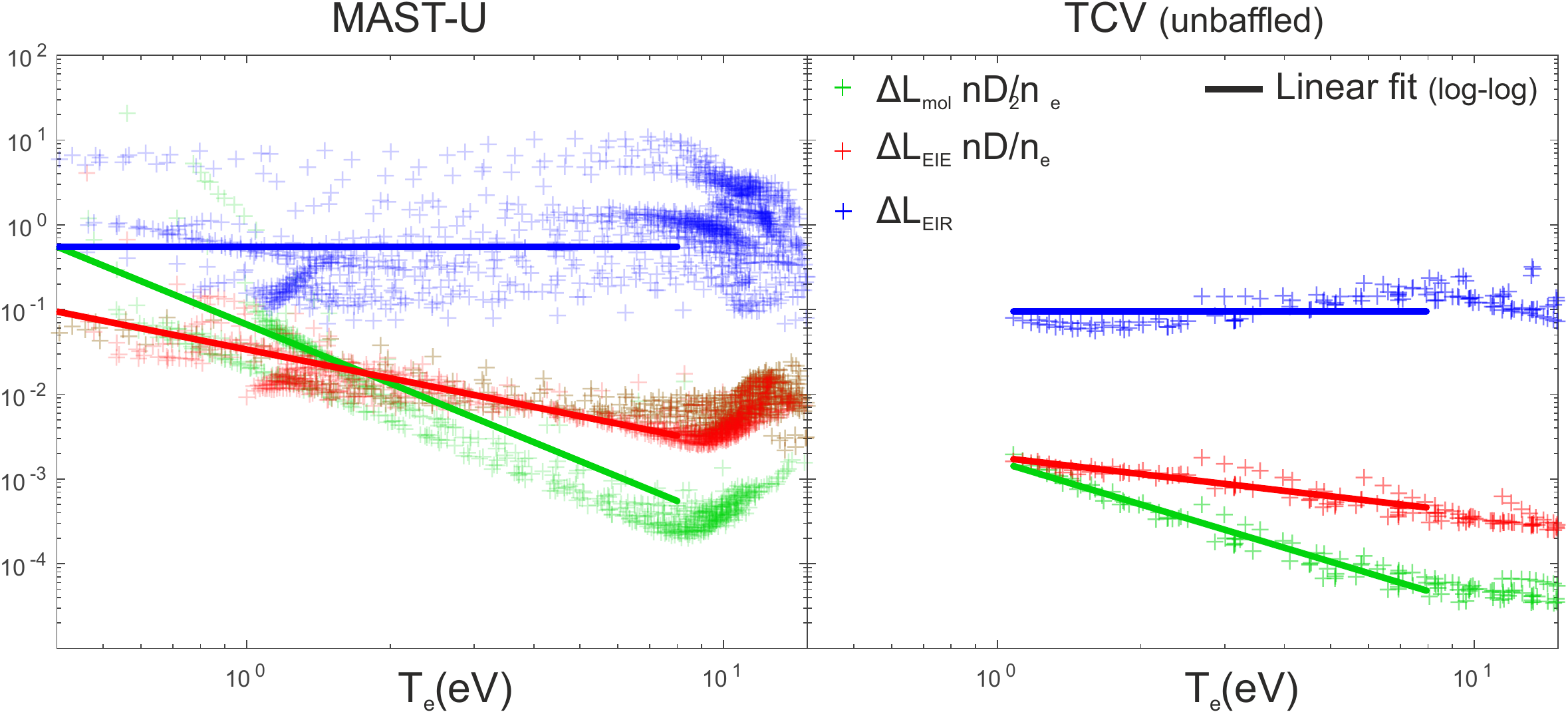}
    \caption{Scalings of $f_{\Delta L n_D/n_e}^{MAST-U} (T_e)$, $f_{\Delta L n_D/n_e}^{TCV} (T_e)$, $f_{\Delta L n_{D_2}/n_e}^{MAST-U}$, $f_{\Delta L n_{D_2}/n_e}^{TCV}$, $f_{\Delta L_{rec}}^{MAST-U}$, $f_{\Delta L_{rec}}^{TCV}$ obtained from SOLPS-ITER \cite{Fil2017,Myatra} using synthetic diagnostics \cite{Verhaegh2019a,Verhaegh2021,Wensing2019} as function of their respective temperatures (excitation temperature for EIE, PMI and recombination temperature for EIR).}
    \label{fig:scalings}
\end{figure}

$f_{\Delta L n_H/n_e}$, $f_{\Delta L_{rec}}$ and $f_{\Delta L n_{D_2}/n_e}$ have been obtained using synthetic diagnostic techniques employed to both TCV as well as MAST-U SOLPS-ITER simulations, which provide us with a synthetic brightness "measurement" of $B_{n\rightarrow2}^{EIE}$, $B_{n\rightarrow2}^{EIR}$ and $B_{n\rightarrow2}^{PMI}$. Using the $D\alpha$ \emph{emissivity} weighted $n_e$ and $T_e$ along each line of sight, $f_{\Delta L n_D/n_e}$, $f_{\Delta L_{rec}}$ and $f_{\Delta L n_{D_2}/n_e}$ can be calculated for every diagnostic chord and every SOLPS-ITER simulation. For MAST-U both the V1 and V2 views have been used to compute these relations (figure \ref{fig:GeomLoS}). Using the obtained database for those parameters for TCV (density scan) and MAST-U (density scan and $N_2$ seeding), a linear fit in logarithmic space \cite{Stangeby2017,Verhaegh2021} is performed through the obtained database for $f_{\Delta L n_H/n_e}$, $f_{\Delta L_{rec}}$ and $f_{\Delta L n_{D_2}/n_e}$ to obtain the used scalings, shown in equation \ref{eq:scalings} and figure \ref{fig:scalings} as function of temperature.

\begin{equation}
\begin{aligned}
    f_{\Delta L n_D/n_e}^{MAST-U} (T_e) &= 10^{-1.4721 - 1.126 \log 10 (T_e)} \text{ m}\\
    f_{\Delta L n_D/n_e}^{TCV} (T_e) &= 10^{-2.7442 - 0.658 \log 10 (T_e)} \text{ m} \\
    f_{\Delta L n_{D_2}/n_e}^{MAST-U} (T_e) &= 10^{-1.176 - 2.3075 \log 10 (T_e)} \text{ m} \\
    f_{\Delta L n_{D_2}/n_e}^{TCV} (T_e) &= 10^{-2.7907 - 1.6965 \log 10 (T_e)} \text{ m} \\
    f_{\Delta L_{rec}}^{MAST-U} &= 10^{-0.2609} \text{ m} \\
    f_{\Delta L_{rec}}^{TCV} &= 10^{-1.0224} \text{ m} \label{eq:scalings}
\end{aligned}
\end{equation}

The emission coefficients were obtained using ADAS \cite{OMullane,Summers2006} ($PEC_{n\rightarrow2}^{EIE}$ \& $PEC_{n\rightarrow2}^{EIR}$) and Yacora (on the Web) \cite{Wunderlich2021,Wunderlich2020,Wuenderlich2016} ($PEC_{n\rightarrow2}^{PMI}$). $PEC_{n\rightarrow2}^{PMI}$ is an effective emission coefficient which accounts for interactions involving $D_2, D_2^+$ and $D^-$: $PEC_{n\rightarrow2}^{PMI} (n_e, T_e) = PEC_{n\rightarrow2}^{D_2} (n_e, T_e) + \frac{n_{D_2^+}}{n_{D_2}} PEC_{n\rightarrow2}^{D_2^+} (n_e, T_e) + \frac{n_D{^-}}{D_2} PEC_{n\rightarrow2}^{D^-} (n_e, T_e)$, where $\frac{n_{D_2^+}}{n_{D_2}}$ and $\frac{n_{D^-}}{n_{D_2}}$ are the ratios between the $D_2^+$ and $D^-$ densities to the $D_2$ density, respectively. Those ratios are modelled as function of $n_e$ and $T_e$ using AMJUEL/Sawada \cite{AMJUEL,Sawada1995,Reiter2018,Verhaegh2021,Verhaegh2021b} (H12 2.2.0c and H11 7.2.0c respectively). However, those AMJUEL rates are specifically derived for hydrogen and not deuterium. Following previous work \cite{Reiter2018}, the hydrogen rates are remapped to the deuterium rates by multiplying them with 0.95 and 0.7 respectively. 

The Fulcher brightness (arbitrary units) is modelled using $\propto n_e^2 f_{\Delta L n_{D_2} / n_e} PEC_{D_2}^{Fulcher} (n_e, T_e)$, where the Fulcher emission coefficient is obtained from AMJUEL/Sawada \cite{AMJUEL,Sawada1995} (H12 2.2.5fl).












\section{Correlation between Fulcher emission and the ionisation source}
\label{ch:fulchemiss_ioni}

\begin{figure}
    \centering
    \includegraphics[width=0.8\linewidth]{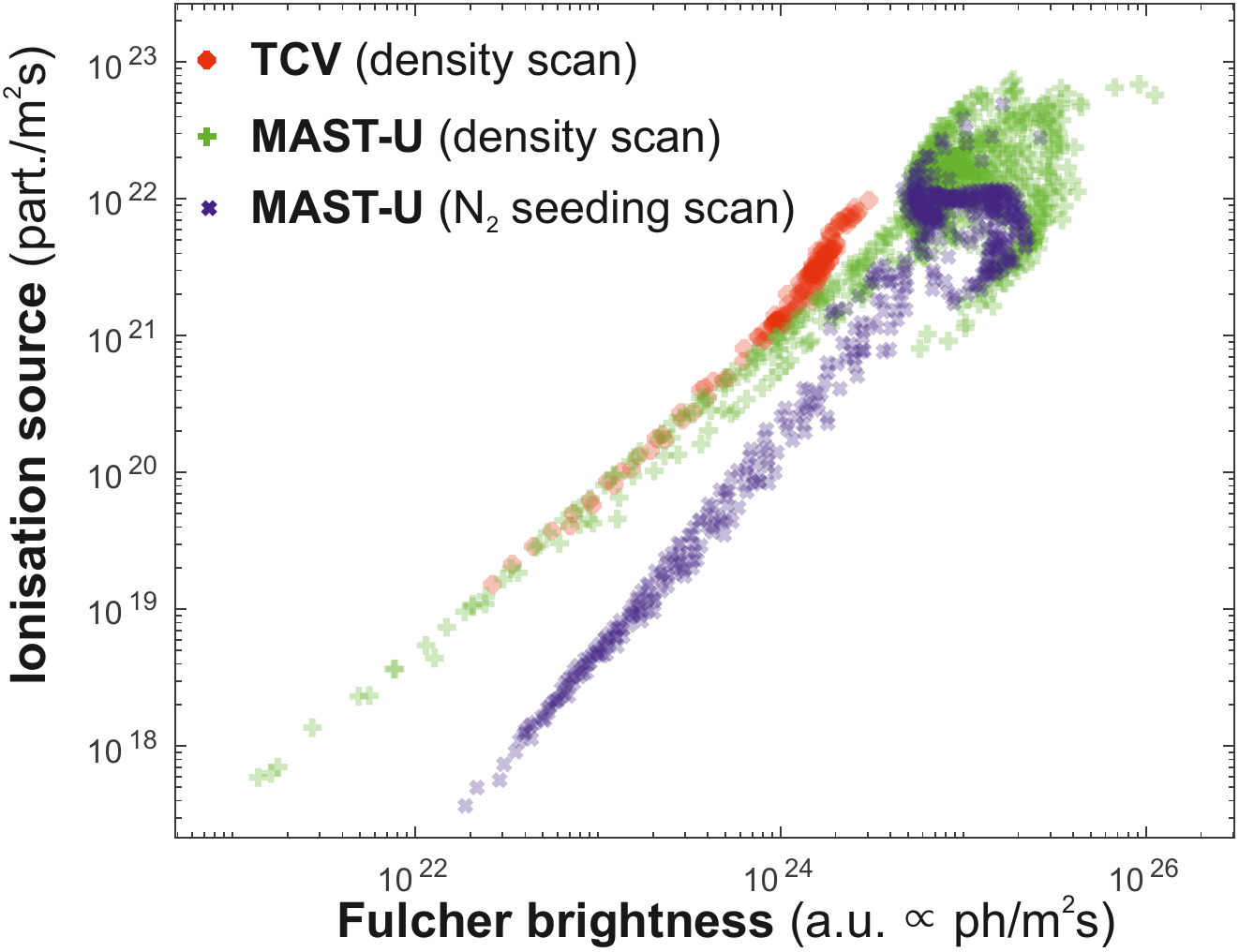}
    \caption{The chordal-integrated (atomic) ionisation source plotted as function of the chordal-integrated Fulcher emissivity (a.u.) using synthetic diagnostics applied to SOLPS-ITER simulations (5 simulations TCV, density scan \cite{Fil2017}; 15 simulations MAST-U (density scan) \cite{Myatra} and 19 simulations MAST-U ($N_2$ seeding scan) \cite{Myatra}. The experimental spectroscopy chord descriptions of TCV (32 lines of sight, before baffles were installed \cite{Verhaegh2017}) and MAST-U (40 lines of sight, figure \ref{fig:GeomLoS}, no interleaving) have been used to determine the result.}
    \label{fig:IonFulcher}
\end{figure}

We investigate the suitability of using the Fulcher emission brightness as a diagnostic for the ionisation region by applying a synthetic diagnostic, for both the line-integrated Fulcher band emission brightness ($\propto ph/m^2/s$) and the atomic ionisation source ($ion/m^2/s$), to SOLPS-ITER simulations using all spectroscopic viewing chords (DSS on TCV \cite{Verhaegh2017} and DMS on MAST-U (figure \ref{fig:GeomLoS})). The two are plotted as function of each other in figure \ref{fig:IonFulcher}, displaying a strong correlation, particularly for decaying levels of the ionisation source (corresponding to lower temperatures ($T_e \ll 5$ eV)). This correlation is found for both MAST-U (density scan and seeded cases) and TCV (density scan). Although the precise correlation is not expected to be the same for different conditions (as it depends on the precise ratio between the neutral and molecular density), a similar correlation is found for a density scan for TCV and MAST-U despite the different plasma conditions; molecular / neutral densities and chordal integrals. During $N_2$ seeding, however, the obtained correlation is different due to lower neutral fractions at lower temperatures compared to the density ramp case \cite{Myatra}. 

One caveat to this analysis lies in the difference in mean free path for atoms and molecules. Although they both become larger than the divertor leg dimension during detached regimes, in attached conditions (for the studied regimes) the neutral atom mean free path is of a similar order to the divertor size while that of molecules is significantly smaller. Therefore, the correlation between the Fulcher emission region and the ionisation region is expected to worsen in higher temperature conditions (observed in figure \ref{fig:IonFulcher}). 

\section{Analysis of EIR contribution to high-n Balmer lines}
\label{ch:highnEIR}

\begin{figure}
    \centering
    \includegraphics[width=\linewidth]{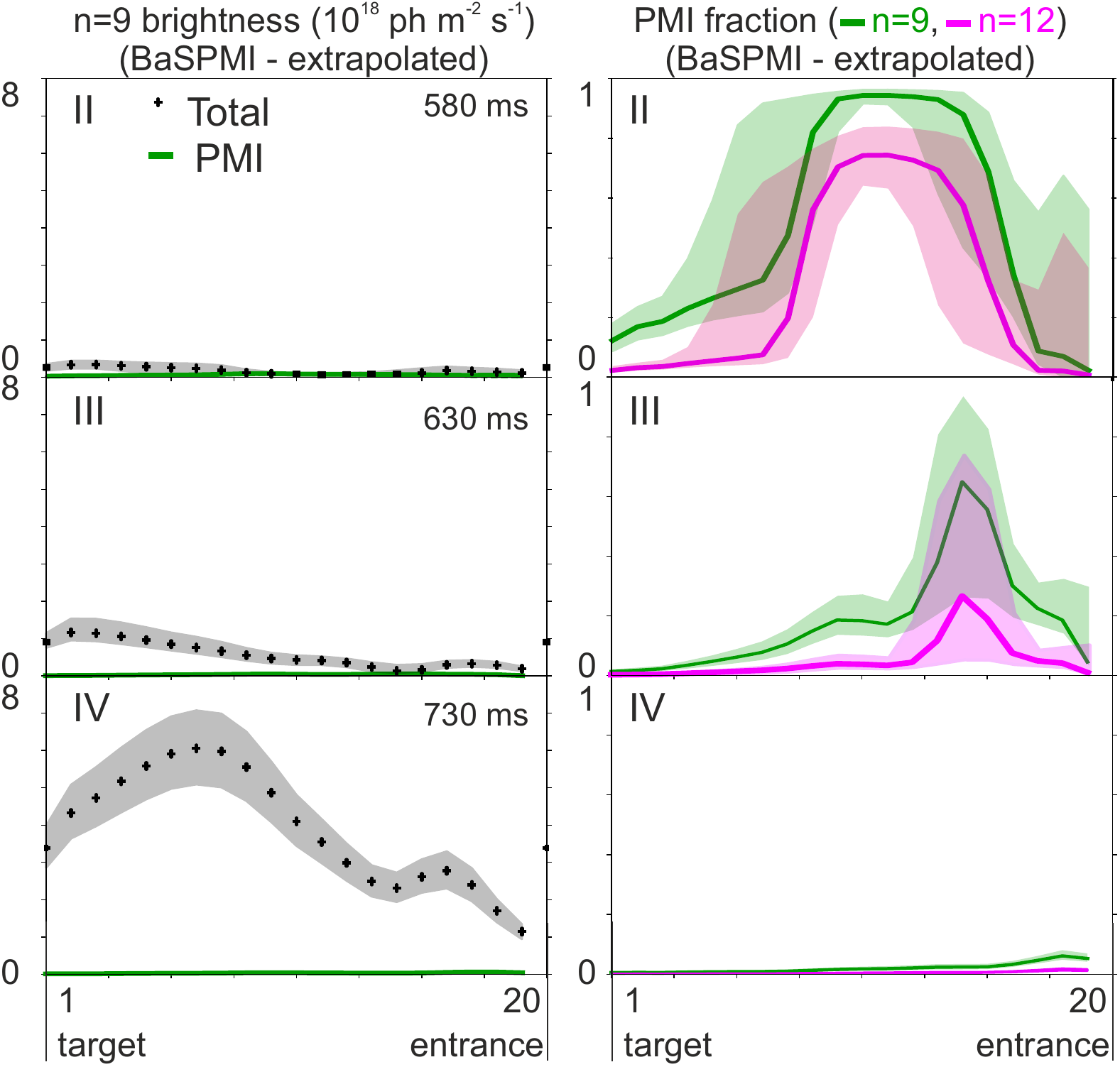}
    \caption{Extrapolated total and plasma-molecule interaction related contribution ('PMI') to the $n=9$ Balmer line emission for \# 45371 during phases II, III, and IV, together with the fraction of the $n=9$ (green) and $n=12$ (magenta) Balmer line emission due to plasma-molecule interactions.}
    \label{fig:n9contamin}
\end{figure}

In section \ref{ch:strong_EIR} the high-n Balmer line analysis was shown, where it is assumed that the emission of the high-n Balmer lines (e.g. $n \geq 9$) is dominated by electron-ion recombination. To challenge this assumption, we post-process the BaSPMI analysis output from \# 45371 to estimate 1) the total high-n Balmer line intensities; 2) the contribution of plasma-molecular interactions to high-n ($n \geq 9$) Balmer lines (figure  \ref{fig:n9contamin}). This indicates that the high-n ($n\geq 9$) Balmer line emission may, at most, include a 10 \% contribution from plasma-molecular interactions during the detachment phase IV. Although the contribution of plasma-molecular interactions to the $n=9$ Balmer line is larger before EIR commences,  the contribution of the $n=9$ Balmer line emission is at most $10^{17} ph m^{-2} s^{-1}$. This corresponds to below the detection threshold of our instrument. 

\section{BaSPMI analysis set-up}
\label{ch:baspmi}

BaSPMI \cite{Verhaegh2021} has been used in this work to separate the $D\alpha$ emission profile into its various atomic and molecular contributions (figure \ref{fig:EmissionTrends}). The BaSPMI analysis is performed using the measured brightnesses shown in figure \ref{fig:EmissionTrends} ($n=3,5,6$ Balmer line emission). In BaSPMI, the ratio between the molecular contributions of $D\alpha$ and $D\beta$ are typically used to separate $D_2^+$ and $D^-$ emission contributions \cite{Verhaegh2021}, which requires $D\beta$ measurements. Since $D\beta$ measurements are not available for this specific discharge, the separation between $D_2^+$ and $D^-$ is modelled using AMJUEL/Sawada rates \cite{AMJUEL,Sawada1995} instead (similar to the emission model used in  \ref{ch:emissmodel}). 

To facilitate these preliminary BaSPMI calculations on MAST-U, various simplifying assumptions are made. First, instead of estimating $\Delta L$ using the width of the ion saturation current profile near the target, an assumed width near the target of 10 cm is used (from the separatrix towards the low-field-side). $\Delta L$ is then computed by calculating the intersections between the line of sight and the flux surfaces corresponding to this width. The lower/upper ranges of $\Delta L$ (68 \% confidence interval) are set to 33 and 130 \% of $\Delta L$, respectively, and are used in the uncertainty propagation.

To employ BaSPMI, the Fulcher band has been used to provide an excitation temperature constraint (section \ref{ch:fulcher_ionisation}). This is achieved by using the trailing end of the Fulcher brightness profile along the various spectroscopic chords, which are normalised with respect to the peak and after the peak of the Fulcher band brightness has left the spectroscopic view, the last known peak intensity is used for the normalisation.

The part of the analysis that obtains an estimate of the recombinative temperature $T_e^R$ had to be altered as the recombinative temperatures on MAST-U seem to be below the ADAS limit of 0.2 eV. As such, no suitable temperatures could be found to explain the measured recombinative emission brightness during conditions where strong recombinative emission occurs. As opposed to filtering out these results from the Monte Carlo sample, $T_e^R$ has been set to 0.2 eV in this regime.

BaSPMI employs an iterative scheme, which uses an initial condition on the contribution of plasma-molecular interactions (PMI) to the medium-n Balmer lines and is then iterated until convergence. By default, the initial condition used is that the contribution of PMI to the medium-n Balmer lines is zero. For MAST-U, due to the strong contributions of PMI to the hydrogen Balmer line emission resulting in a reduction of the $6 \rightarrow 2 / 5 \rightarrow 2$ Balmer line ratio, the initial conditions of the PMI contribution to the medium-n Balmer lines had to be increased to obtain reasonable solutions consistent with the various other diagnostics (DTS estimates, Fulcher brightness information, other information); otherwise the temperature would be grossly overestimated by BaSPMI. The BaSPMI results with the modified initial conditions are verified against an independent fully Bayesian analysis which does not require any initial conditions in \ref{ch:bayspmi}.


An important caveat to the presented analysis is the assumption that the plasma is optically thin. This could not be verified experimentally without divertor VUV spectroscopy. If the plasma is not optically thin, it could greatly impact the analysis results, as explained in \cite{Verhaegh2021a,Verhaegh2021}, and could alter both the reaction rates as well as the emission coefficients \cite{Lomanowski2020,Lomanowski2019}. Although photon opacity may occur in MAST-U conditions \cite{Verhaegh2021b,Soukhanovskii2022}, it is expected to occur at much higher powers ($P_{sep} = 2.5 - 5$ MW compared to 0.5 MW (this work)) and electron densities ($n_e > 10^{20} m^{-3}$ compared to $n_e \sim 10^{19} m^{-3}$ in this work) \cite{Soukhanovskii2022}. However, this will require further monitoring in the future.

\subsection{Bayesian to BaSPMI comparison}
\label{ch:bayspmi}

To verify these modifications to the BaSPMI analysis, a stand-alone fully Bayesian analysis methodology has been used as well using the same emission model as BaSPMI and using priors for the excitation emission temperature based on the Fulcher band; $n_e$ based on Stark broadening; $\Delta L$ based on the BaSPMI input for $\Delta L$ same constraints; a constraint that the excitation emission temperature should be larger than the recombination emission temperature \cite{Verhaegh2019a}. Uncertainties in the emission coefficients are not employed in the Bayesian analysis, but are used in BaSPMI (10 \% for atomic PECs and 20 \% for molecular PECs). The measurements used in this Bayesian approach are the $n=3, 5, 6$ Balmer line brightnesses. The Bayesian analysis results in a posterior distribution and rejection sampling is applied to obtain the same sets of output data similar to BaPSMI. To keep the number of free parameters at 6, the Bayesian implementation only includes contributions from $D_2^+$, electron-impact excitation and electron-ion recombination (e.g. not $D_2$ and $D^-$). To reduce the computational effort needed, the sample size is reduced to 500 in the Bayesian analysis, whereas this is 30000 in BaSPMI (effective sample size (Kish's rule) of 500 and 2500-8000, respectively).  

\begin{figure}
    \centering
    \includegraphics[width=\linewidth]{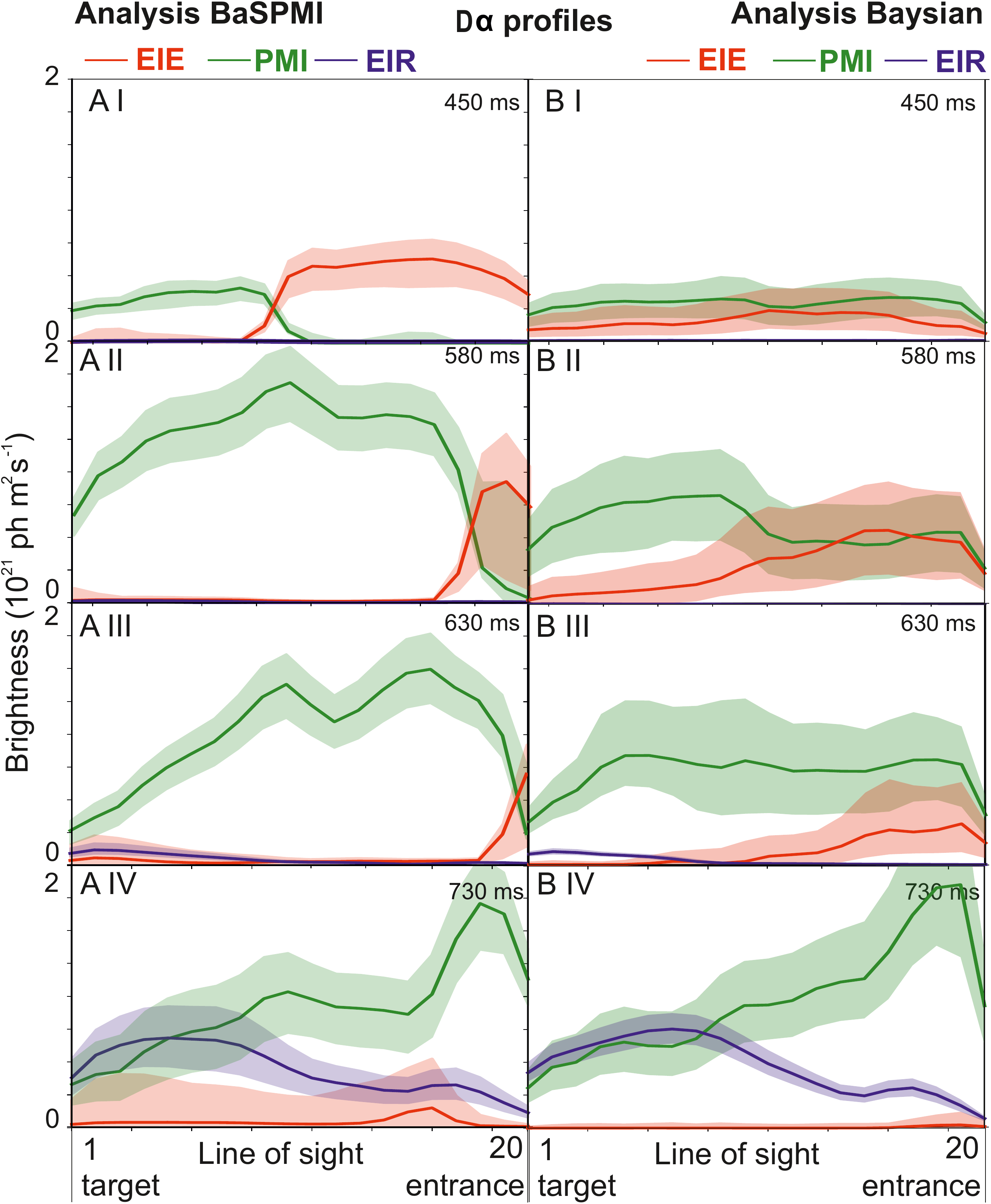}
    \caption{Comparison of figure \ref{fig:EmissionTrends} A I-IV obtained using BaSPMI and B I-IV obtained using an alternative Bayesian approach for \# 45371.}
    \label{fig:bayspmi}
\end{figure}

This is then analysed in the same way as BaSPMI to obtain the same parameters as those shown in figure \ref{fig:EmissionTrends} A I-IV, which is shown in figure \ref{fig:bayspmi} B I-IV. The obtained result is qualitatively in agreement with BaSPMI. Modelling the Bayesian results back to the hydrogen emission brightnesses results in a reasonable agreement to the measured data - implying that the Bayesian analysis results are indeed likely outputs from the data. 

There are, however, reasons for the differences between BaSPMI and the Bayesian analysis approach. First, the BaSPMI technique assumes that $F_{rec}$ \cite{Verhaegh2019a} - the fraction of recombinative emission compared to the EIR + EIE Balmer line emission - cannot decrease during a discharge - constraining the obtained results further; which is something that cannot easily be implemented in a full Bayesian approach. Secondly, the Bayesian analysis will struggle more with the fact that ADAS data is (currently) only available until 0.2 eV - therefore it cannot describe the EIR portion of the emission accurately ($B_{n\rightarrow2}^{EIR, Bayesian} \approx n_e^2 \Delta L PEC_{n\rightarrow2}^{EIR} (n_e, T_e^R)$). This is less of a problem for BaSPMI as the EIR portion of the emission is characterised using the fraction of medium-n Balmer line emission due to plasma-molecule interactions ($f_{n\rightarrow2}^{PMI}$) and the fraction of the atomic portion of the medium-n Balmer line emission due to EIR ($F_{rec}^{n\rightarrow2}$) - e.g. $B_{n\rightarrow2}^{EIR, BaSPMI} = (1-f_{n\rightarrow2}^{mol}) \times F_{rec}^{n\rightarrow2} B_{n\rightarrow2}$. The Bayesian analysis compensates for this by increasing the $n_e$ to $3 \times 10^{19} m^{-3}$ (clearly too high when compared with $<10^{19} m^{-3}$ values from DTS as well as $<2 \times 10^{19} m^{-3}$ values from Stark broadening) in the recombining regime while keeping $T_e^R$ at 0.3 eV.

\end{document}